\documentclass[floatfix,a4paper,aps,prd,twocolumn]{revtex4}

\usepackage{epsfig}
\usepackage{axodraw}
\usepackage{amsmath}
\usepackage{amssymb}
\usepackage{amsfonts}
\usepackage{graphicx}
\usepackage{graphics,subfigure}
\usepackage{axodraw}
\usepackage{float}
\usepackage{booktabs}
\usepackage{geometry}
\usepackage{rotating}

\newcommand{\newc}{\newcommand}
\newc{\ba}{\begin{array}}
\newc{\ea}{\end{array}}
\newc{\bea}{\begin{eqnarray}}
\newc{\eea}{\end{eqnarray}}
\newc{\beastar}{\begin{eqnarray*}}
\newc{\eeastar}{\end{eqnarray*}}
\newc{\beq}{\begin{equation}}
\newc{\eeq}{\end{equation}}
\newc{\bestar}{\begin{equation*}}
\newc{\eestar}{\end{equation*}}
\newc{\ben}{\begin{enumerate}}
\newc{\een}{\end{enumerate}}
\newc{\bi}{\begin{itemize}}
\newc{\ei}{\end{itemize}}
\newc{\mymed}{\vspace{0.14cm}}

\newc{\lam}{\lambda}
\newc{\lamp}{\lambda^\prime}
\newc{\lampp}{\lambda^{\prime\prime}}
\newc{\Lam}{\Lambda}
\newc{\BLam}{{\mathbf{\Lam}}}
\newc{\eps}{\epsilon}
\newc{\kap}{\kappa}

\newc{\ra}{\rightarrow}
\newc{\ovl}{\overline}
\newc{\lsim}{\stackrel{<}{\sim}}

\newc{\itRahmen}[2]{
  \begin{center}\fbox{\parbox{#1 cm}{\it #2}}
\end{center}}
\newc{\del}{\partial}
\newc{\veva}{\langle H_1\rangle}
\newc{\vevb}{\langle H_2\rangle}
\newc{\onehalf}{\textstyle \frac{1}{2} \displaystyle}
\newc{\onethird}{\textstyle \frac{1}{3} \displaystyle}
\newc{\mzero}{M_0}
\newc{\mhalf}{{M_{1/2}}}
\newc{\tanb}{\tan\beta}
\newc{\Psix}{{\mathrm{P}_{\!6}}}
\newc{\nPsix}{{\not\!\Psix}}
\newc{\nPsixU}{{\s{\mathrm P}_{\!6}}}

\newc{\azero}{A_0}
\newc{\neutralino}{\tilde\chi^0}
\newc{\selectron}{\tilde e}
\newc{\stau}{\tilde\tau}
\newc{\smuon}{\tilde\mu}
\newc{\higgs}{h^0}
\newc{\sgnmu}{\textrm{sgn}(\mu)}
\newc{\gev}{\mbox{~GeV}}
\newc{\tev}{\mbox{~TeV}}
\newc{\gsim}{\stackrel{>}{\sim}}

\newbox\charbox
\newbox\slabox
\def\s#1{{      % Feynman slash
    \setbox\charbox=\hbox{$#1$}
    \setbox\slabox=\hbox{$/$}
    \dimen\charbox=\ht\slabox
    \advance\dimen\charbox by -\dp\slabox
    \advance\dimen\charbox by -\ht\charbox
    \advance\dimen\charbox by \dp\charbox
    \divide\dimen\charbox by 2
    \raise-\dimen\charbox\hbox to \wd\charbox{\hss/\hss}
    \llap{$#1$}
}}

\newc{\onegraph}[4]{%
  \unitlength=1in
  \begin{picture}(3,2.3)
    \put(-0.6,0){\epsfig{file=#1.eps, width=3.7in}}
    \put(0.33,0.48){\epsfig{file=#12.eps, width=2.509in}}
    \put(2.9,1.1){\rotatebox{90}{$#2$}}
    \put(1.3,0.2){\makebox(0,0){$#3$}}
    \put(0.05,1.2){\rotatebox{90}{$#4$}}
  \end{picture}
}

\newc{\ssup}{\tilde{u}}
\newc{\ssdown}{\tilde{d}}
\newc{\ssstrange}{\tilde{s}}
\newc{\sscharm}{\tilde{c}}
\newc{\sstop}{\tilde{t}}
\newc{\ssbottom}{\tilde{b}}

\newc{\sse}{\tilde{e}}
\newc{\ssmu}{\tilde{\mu}}
\newc{\sstau}{\tilde{\tau}}
\newc{\ssnue}{\tilde{\nu}_{e}}
\newc{\ssnumu}{{\tilde{\nu}_{\mu}}}
\newc{\ssnutau}{{\tilde{\nu}_{\tau}}}
\newc{\ssbnue}{\bar{\tilde{\nu}}_{e}}
\newc{\ssbnumu}{\bar{\tilde{\nu}}_{\mu}}
\newc{\ssbnutau}{\bar{\tilde{\nu}}_{\tau}}

\newc{\neut}{{\tilde\chi}^0}
\newc{\charge}{{\tilde\chi}}
\newc{\glu}{\tilde{g}}

\newc{\Higgs}{H^0}
\newc{\Azero}{A^0}

\newc{\nue}{\nu_e}
\newc{\numu}{\nu_{\mu}}
\newc{\nutau}{\nu_{\tau}}
\newc{\bnue}{\bar{\nu_e}}
\newc{\bnumu}{\bar{\nu_{\mu}}}
\newc{\bnutau}{\bar{\nu_{\tau}}}
\newc{\Br}{\mathrm{Br}}

\begin{document}
  
%\begin{flushright}
%\hspace{12cm} a
%\end{flushright}
  
\title{Mass Spectrum in R-Parity Violating minimal Supergravity and Benchmark Points\footnote{Preprint number: DAMTP-2006-76}}
  
\author{B.~C.~Allanach\footnote{\,allanach@cern.ch}}

\affiliation{DAMTP, University of Cambridge, Cambridge, UK}

\author{M.~A.~Bernhardt\footnote{{\,markus@th.physik.uni-bonn.de}}}

\affiliation{Physics Institute, University of Bonn, Bonn, Germany}

\author{H.~K.~Dreiner\footnote{\,dreiner@th.physik.uni-bonn.de}}

\affiliation{Physics Institute, University of Bonn, Bonn, Germany}

\author{C.~H.~Kom\footnote{\,c.kom@damtp.cam.ac.uk}}

\affiliation{DAMTP,\phantom{.}\!\! University of Cambridge, Cambridge, UK}

\author{P.~Richardson\footnote{\,Peter.Richardson@durham.ac.uk}}

\affiliation{IPPP, University of Durham, Durham, UK}

\begin{abstract}
  We investigate in detail the low-energy spectrum of the R-parity
  violating minimal supergravity model using the computer program
  \texttt{SOFTSUSY}. We impose the experimental constraints from the
  measurement of the anomalous magnetic moment of the muon,
  $(g-2)_\mu$, the decay $b\ra s\gamma$ as well as the mass bounds
  from direct searches at colliders, in particular on the Higgs boson
  and the lightest chargino. We also include a new calculation for the
  R parity violating contribution to $\Br(B_s\ra \mu^+\mu^-)$.
  We then focus on cases where the lightest neutralino is {\em not} the
  lightest supersymmetric particle (LSP). In this region of parameter
  space either the lightest scalar tau (stau) or the scalar tau neutrino
  (tau sneutrino)
  is the LSP. We suggest four benchmark points with typical spectra
  and novel collider signatures for detailed phenomenological
  analysis and simulation by the LHC collaborations.
\end{abstract}

\maketitle

\section{Introduction}

The most widely discussed solution to the gauge hierarchy problem
\cite{hierarchy} of the Standard Model of particle physics (SM) \cite{
  SM} is the supersymmetric extension of the SM (SSM) \cite{
  Wess:1974tw}. However, no supersymmetric particle has been observed
to-date. Thus if it exists supersymmetry (SUSY) must be broken and the
mass scale should be of order the TeV energy scale, in order to
maintain the solution to the hierarchy problem \cite{drees}. The TeV
energy realm will be probed at the LHC starting in 2007 with the
search for SUSY of paramount interest \cite{susy-lhc}.

\mymed

The general SSM renormalisable superpotential with minimal particle
content is given by \cite{ssm-superpot}
\bea
W_{\mathrm{SSM}}&=& W_{\Psix}+W_{\nPsix}\,,
\label{superpot} \\[1.5mm]
W_{\Psix}&=&\eps_{ab}\left[(\mathbf{Y_E})_{ij}L_i^aH_1^b
  \ovl{E}_j + (\mathbf{Y_D})_{ij}Q_i^{ax}H_1^b\ovl{D}_{jx}
\right.  \notag \\ & & 
\left.+(\mathbf{Y_U})_{ij}Q_i^{ax}H_2^b\ovl{U}_{jx} + \mu
  H_1^aH_2^b\right],\label{P6-superpot} \\[1.5mm]
W_{\nPsix} & = & \eps_{ab}\left[\frac{1}{2} \lam_{ijk} L_i^aL_j^b
\ovl{E}_k + \lam'_{ijk}L_i^aQ_j^{bx}\ovl{D}_{kx}\right]\notag \\
&&%\hspace{-0.7cm}
+\eps_{ab}\kap^i  L_i^aH_2^b
+\frac{1}{2}\eps_{xyz}{\lam}''_{ijk}
\ovl{U}_i^{\,x} \ovl{D}_j^{\,y} \ovl{D}_k^{\,z} \,.
\label{P6v-superpot}
\eea
Here we have employed the notation of Ref.~\cite{Dreiner:1997uz}.
We have split the superpotential into two sets of interactions,
for reasons which we shall explain shortly. 

\mymed

If simultaneously present, the baryon and lepton number violating
interactions in Eq.~(\ref{P6v-superpot}) lead to rapid proton decay
\cite{proton-decay}.  Therefore the SSM must be augmented by a
symmetry. The most widely studied scenario is R-parity
\cite{Farrar:1978xj}, for which $W_{\nPsix}=0$.  However, it suffers
from dangerous dimension-five proton-decay operators
\cite{ssm-superpot}. This is solved by proton hexality, $\Psix$, an
anomaly-free $\mathbf{Z} _6$ discrete gauge symmetry
\cite{Dreiner:2005rd}. The renormalisable superpotential terms are
equivalent to conserved R-parity, namely $W_\Psix$, but the
dimension-five proton-decay operators are forbidden.  So we consider
this the more appropriate choice and have correspondingly labelled the
superpotential in Eqs.~(\ref{superpot},\ref{P6-superpot}). The $\Psix
$-SSM is conventionally denoted the minimal supersymmetric Standard
Model (MSSM).

\mymed

Baryon triality, $B_3$ is an anomaly-free $\mathbf{Z}_3$ discrete
gauge symmetry \cite{Ibanez:1991hv}, which prohibits the $\ovl{U}\ovl
{D}\ovl {D} $ interactions, thereby stabilising the proton. It is thus
theoretically equally well motivated to $\Psix$. A third possibility
to stabilise the proton is lepton parity \cite{ Ibanez:1991hv}, which
prohibits the lepton-number violating interactions in
Eq.~(\ref{P6v-superpot}). However, lepton parity is anomalous
\cite{Ibanez:1991hv}. As the alternative to the $\Psix$-MSSM we shall
focus on the phenomenological analysis of baryon triality at 
colliders. However, we shall also consider some implications of lepton
parity.

\mymed

The $\Psix$-MSSM mass spectrum and couplings have been studied in
great detail \cite{Ibanez:1983wi,Alvarez-Gaume:1983gj,
Ibanez:1983di}. However, there are two reasons why it is highly
desirable to perform analyses of the \textit{general} SSM at
colliders. First, to make sure that no possibility to discover SUSY is
left untried. If we focus our endeavours on one specific form of SUSY,
due to theoretical prejudice for example, then we might initially
entirely miss the discovery. Second, SUSY is not a simple extension of
the SM, like for example right-handed neutrinos, which does not affect
most parts of the model. In the SSM, the particle spectrum is more
than doubled. If the SSM is to provide a solution to the technical
hierarchy problem, we expect all of the new particles to have masses
$\lsim{\mathcal O}(2\tev)$. Thus if low-energy SUSY exists we should
expect many new processes to be kinematically accessible at the
LHC\@. When considering a specific SUSY process as a signature, many
other SUSY processes will act as background and must be taken into
account \cite{Hinchliffe:1996iu}.

\mymed

Thus in order to perform simulations of signatures involving the
complete supersymmetric spectrum in preparation for the LHC, we 
restrict ourselves to specific points of the $\Psix$-minimal supergravity ($\Psix$-mSUGRA) parameter
space. Over the years several such example points have been proposed
\cite{Hinchliffe:1996iu,benchmark,Allanach:2002nj}. Some of these 
``benchmark points'' have been excluded by searches at LEP, for
example the points 3 and 5 in Ref.~\cite{Hinchliffe:1996iu}. Some
points are strongly motivated by the resulting cosmological relic
density \cite{dark-matter-bench}. A set of benchmark points and
parameter lines in the mSUGRA parameter space have been agreed upon,
the Snowmass Points and Slopes (SPS) \cite{Allanach:2002nj}. The
resulting spectra have been compared for various numerical RGE
integration codes \cite{Azuelos:2002qw, Allanach:2003jw} and have been
investigated both in the context of the LHC and the ILC\@. It is our
purpose to facilitate the extension of this work to the $\not\!
\Psix$-MSSM\@. As a first step, we shall investigate the mass spectrum
and couplings in some detail. For further more detailed experimental
studies we also propose a set of benchmark points.

\mymed

In Sect.~\ref{p6-sect} we briefly review the most widely studied
model, namely the $\Psix$-MSSM embedded in supergravity \cite{
  Freedman:1976xh,Nilles:1983ge}. In order to throw a wider net in our
search for SUSY, we extend this discussion to the $\nPsixU$-MSSM in
Sect.~\ref{sec:p6v}, emphasising the main differences to the $\Psix$
conserving case and examining bounds on the $\nPsixU$ couplings. In
this first study we sometimes will focus on the simplest model of no-scale
supergravity \cite{Ellis:1983sf}.  We examine constraints on $\nPsixU$
mSUGRA parameter space arising from precision measurements in
section~\ref{sec:space} and the sparticle spectrum in
section~\ref{sec:spec}.  We present the definition of the proposed
mSUGRA benchmark points in section~\ref{sec:bench}, where we
also detail the spectrum and sparticle decays. We summarise and
conclude in section~\ref{sec:conc}, following with expressions for the
$\nPsixU$ contribution to $\text{Br}(B_{q_k}\ra e^+_m e^-_l)$
decays, a generalisation of results in the literature in
Appendix~\ref{app:one}.  In this paper we do not consider alternative
mechanisms to communicate to the visible sector such as gauge or
gaugino mediated SUSY breaking \cite{NoGMSB} or anomaly mediated SUSY
breaking \cite{NoAMSB}.

\section{The $\Psix$-MSSM \label{p6-sect}}

The $\Psix$-MSSM has 124 free parameters \cite{Haber:1997if}, and the
$\nPsixU$-MSSM more than 200. Such an extensive parameter space is
intractable for a systematic phenomenological analysis at colliders.
It is thus mandatory to consider simpler models, which represent the
variety in the SSM\@. Hereby both observational hints
\cite{Amaldi:1991cn} and theoretical considerations \cite{msugra} have
been employed as guiding tools.

\mymed

Within the SSM, the unification of the three SM gauge couplings at a
scale $M_X=\mathcal{O}(10^{16}\; \textrm{GeV})$ \cite{Amaldi:1991cn}
indicates the embedding in a unified model. The most widely studied
and the simplest of these models is mSUGRA with
conserved $\Psix$ \cite{msugra} and radiative electroweak symmetry
breaking \cite{Ibanez:1982fr}. In this case, we are left with five
free parameters
\beq
\mzero\,,\, \mhalf\,,\, \azero\,,\,\tanb\,,\, \sgnmu\,
\label{mSUGRAparam}
\eeq
where the first three impose boundary conditions on the soft-SUSY breaking parameters at $M_X$, while the other two are boundary conditions at $M_Z$. Here $\mzero$ is the universal, \textit{i.e.} flavour
independent, soft breaking scalar mass. We thus have for example
\beq
m_{{\tilde \nu}_L}\!=m_{{\tilde \ell}_{L,R}}\!
=m_{{\tilde q}_{L,R}}=m_{H_{1,2}}=\mzero,\;\;\;\;\; @M_X \,,\!\!\!
\eeq
where $m_{{\tilde \nu}_L}$ denotes the sneutrino mass, $m_{{
\tilde \ell}_{L,R}}\!$ denote the $SU(2)$-doublet (L) and $SU(2)$
singlet (R) charged slepton masses, and $m_{{\tilde q}_ {L,R}}\!$
denote the squark masses. $m_{H_{1,2} }$ are the soft breaking Higgs
boson masses. $\mhalf$ denotes the universal gaugino soft breaking
mass and we have for the bino ($M_1$), the $SU(2)$-wino ($M_2$) and
the gluino mass ($M_3$), respectively \cite{ftnt1}
\beq
M_1 =M_2=M_3=\mhalf,\qquad @ M_X\,.
\label{univ-gaugino}
\eeq
$\azero$ is the soft breaking universal trilinear scalar interaction
\cite{Nilles:1982dy}. $\tan\beta=\frac{\upsilon_2}{\upsilon_1}$ is the
ratio of the vacuum expectation values of the two Higgs doublets $H_{1
,2}$ and $\sgnmu$ is the sign of the Higgs mixing parameter of
Eq.~(\ref{P6-superpot}). The absolute value of $\mu$ as well as a
possible bilinear scalar interaction $B_0$ are fixed by the
electroweak symmetry breaking conditions at the SUSY breaking scale
\cite{Ibanez:1982fr}.

\mymed

\begin{figure}[t!]
      \scalebox{.8}{\include{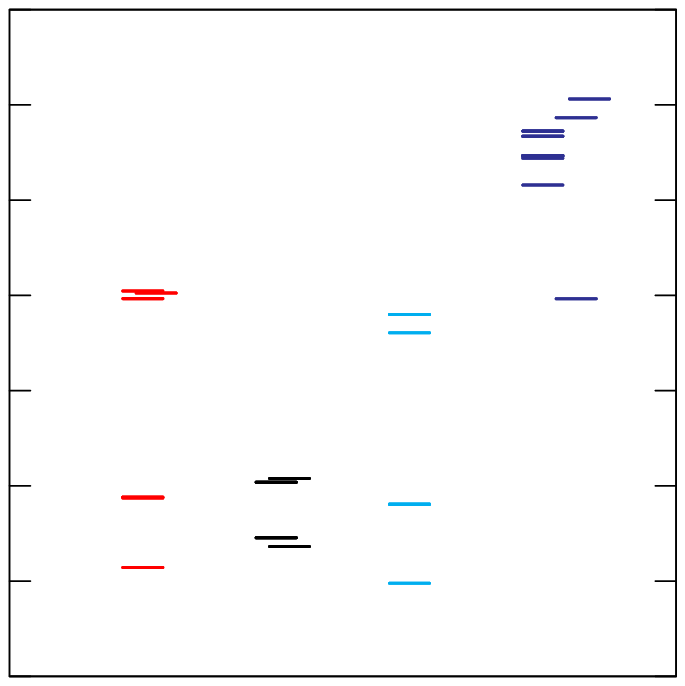}}
      \caption{The SPS1a supersymmetric spectrum. The $\Psix$-mSUGRA
        parameter values are $\mzero=100\gev\,,\,\mhalf=250\gev\,,\,
        \azero=-100\gev\,,\, \tanb=10\,,\,\sgnmu=+1\,$. The up, down, charm and
        strange squarks are quasi-degenerate in mass and are labelled
        by ${\tilde q}_{L,R}$. }
	\label{fig:sps1a}
\end{figure}

Given the five parameters in Eq.~(\ref{mSUGRAparam}), we can
compute the full supersymmetric spectrum and couplings at the
electroweak scale through the renormalisation group equations
(RGEs). This has been studied in great detail
\cite{Ibanez:1983wi,Alvarez-Gaume:1983gj,Ibanez:1983di}. An example
spectrum is shown in Fig.~\ref{fig:sps1a} for the SPS1a parameter set
[$\mzero=100\gev\,,$ $\mhalf=250\gev\,,\,\azero=-100\gev\,,\,\tanb =
10\,,\,\sgnmu=+1\,$] \cite{ Allanach:2002nj}. In the left column
(red), we have the neutral scalars ($h^0,H^0,A^0,\tilde\nu_{Li}$) as
well as the charged Higgs bosons ($H^\pm$). In the second column
(black), we have the charged slepton masses. In the third column
(light blue), we show the neutralinos ($\tilde\chi^0_{ i=1,2,3,4}$) and
charginos ($\tilde\chi^\pm_{1,2} $) and in the fourth column (marine blue)
the squarks ($\tilde u,\tilde d,\tilde c,\tilde s,\tilde t,\tilde b$)
and gluinos $\tilde g$.

\mymed

\begin{figure}[t!]
\begin{center}
\unitlength=1in
\begin{picture}(3,2.3)
	  \put(-0.6,0){\epsfig{file=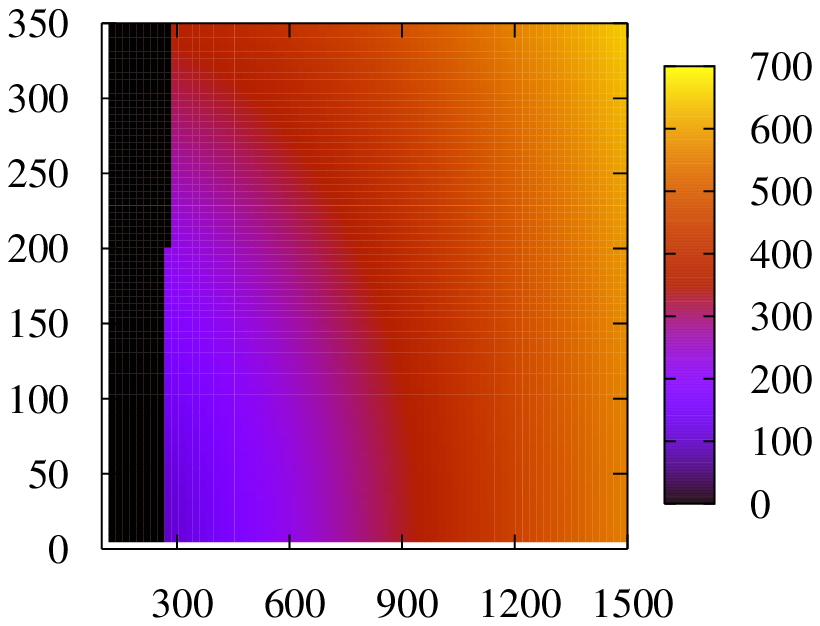, width=3.7in}}
	  \put(0.33,0.48){\epsfig{file=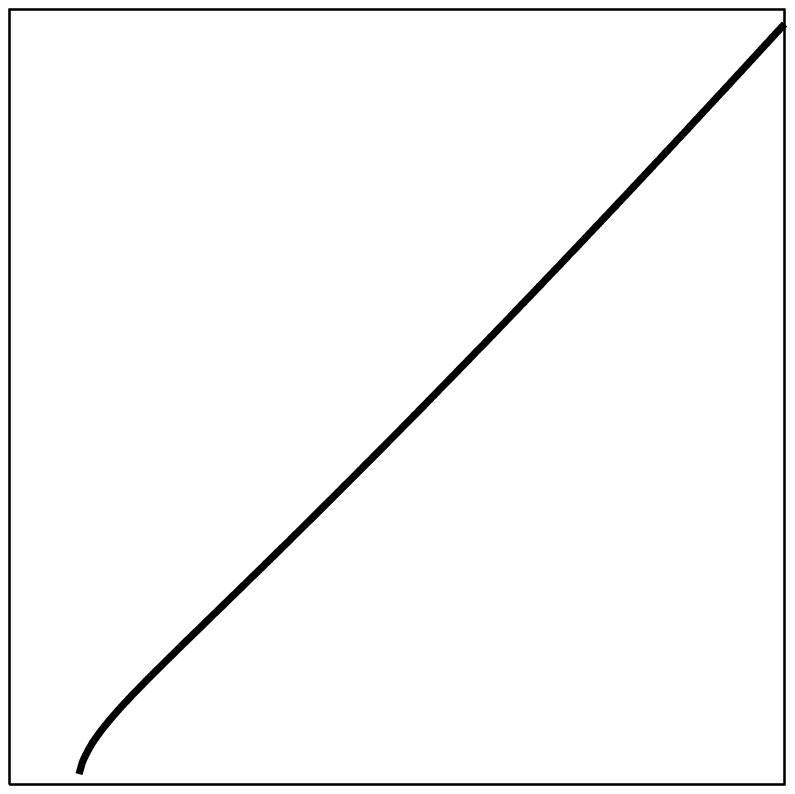, width=2.509in}}
	  \put(2.7,1.1){\rotatebox{90}{$m_{\stau}$ [GeV]}}
	  \put(1.3,0.2){\makebox(0,0){$\mhalf$ [GeV]}}
	  \put(0.0,1.2){\rotatebox{90}{$\mzero$ [GeV]}}
	  \put(1.5,1.0){\makebox(0,0){\White{$\boldsymbol{\stau}$\bf{--LSP}}}}
  \put(1.1,1.6){\makebox(0,0){\White{$\boldsymbol{\neutralino}$\bf{--LSP}}}}
\end{picture}
\end{center}
\caption{Parameter space around SPS1a.
  The bar on the right displays the lightest stau
  mass. $\azero=-100\gev$, $\tan \beta=10$, $\sgnmu=+1$.  The
  blackened out region to the left is excluded due to the LEP bounds
  and tachyons. The black contour distinguishes between areas with
  $\neutralino$-LSP and $\stau$-LSP.}
\label{fig:triangle}
\end{figure}

We next discuss some typical features of this spectrum. First, the
lightest supersymmetric particle (LSP) is the lightest neutralino.
The SPS1a parameter values have been specifically chosen to ensure
this.  Since the LSP is stable in the $\Psix$-MSSM, it must be
electrically and colour neutral for cosmological reasons
\cite{Ellis:1983ew}, and a stable sneutrino LSP is experimentally
excluded \cite{Hebbeker:1999pi}. In $\Psix$-mSUGRA, the regions of
parameter space where the $\neutralino_1$ is not the LSP are thus
excluded from further consideration.  This is important for what
follows since in $\nPsixU$-mSUGRA this region is re-opened. In
Fig.~\ref{fig:triangle}, we have plotted the nature of the LSP in the
$\mzero$-$\mhalf$ plane around the SPS1a point. Most of the $\tilde\chi0-$LSP 
parameter space leads to dark matter relic densities that are too high 
compared with the WMAP3 constraints if $P_6$ is conserved. Since we 
will consider the $\nPsixU$ case where the LSP decays and does not 
constitute dark matter, we do not plot the constraint on Fig.~\ref{fig:triangle}.
The blackened out region at smaller values of $\mhalf$ is excluded 
due to the LEP2 Higgs exclusion bound of $m_h> 114.4 \gev$ at 
$95\%\, \mathrm{C.L.}$ \cite{Barate:2003sz}, and/or the requirement of 
the absence of tachyons in order to obtain a valid electroweak vacuum. 
Note that, anticipating a 3 GeV error in \texttt{SOFTSUSY}'s prediction of $m_h$,
we have imposed a lower bound of 111.4 GeV in the figure. In large
regions of parameter space the $\neutralino_1$ is indeed the LSP.
Separated from this by the black contour, for low values of $\mzero$
and large values of $\mhalf$ the lightest $\stau$ is the LSP\@. For
later use, we also display the stau mass in the figure.

\mymed

Second, in Fig.~\ref{fig:sps1a} it can be seen that the squarks and
gluinos have masses much larger than $\mzero$ and are much heavier
than the sneutrinos and charged sleptons. This is because the strong
interaction dominates the RGE running of the former. For $\mzero\gg
\mhalf$ the slepton masses can be of order the squark and gluino
masses, \textit{cf.} SPS2 \cite{Allanach:2002nj}. Third, one stop, the
one which is predominantly $\tilde t_R$, is significantly lighter than
the other squarks since the stop mixing $\sin \theta_t$ is of order 
$m_t/M_{SUSY}$, whereas $M_{SUSY}$ is the SUSY breaking scale. Similarly the
stau, which is dominantly $\tilde\tau_R$, is 
the lightest charged slepton, although the relative effect is much
smaller, since $m_\tau\ll\mzero$. Since the sleptons themselves are
relatively light the $\tilde \tau_R$ is the next-to-lightest
supersymmetric particle (NLSP). Fourth, the lightest Higgs boson is
significantly lighter than the other Higgs bosons.

\mymed

For SPS1a and other typical mSUGRA points the largest supersymmetric
production cross sections at the LHC are for gluinos and squarks
\cite{LHCXsect}. Usually, squarks and gluinos cascade decay via
intermediate SUSY states to the LSP, leading eventually to a large
missing $p_T$ signature of the $\Psix$-SSM.

\mymed

In order to prepare for the LHC, it is important, given a
supersymmetric spectrum, to analyse the resulting production processes
and decays in great detail. In particular, this involves the full
detector simulation for each signature. Even with the highly reduced
parameter space of $\Psix$-mSUGRA, this is beyond present
capabilities. There is a special mSUGRA model with even fewer
parameters: the so-called no-scale supergravity model
\cite{no-scale}, where we have as additional conditions
\beq 
\azero=\mzero=0\,,\qquad @\,M_X\,. \label{noscale} 
\eeq
This is however experimentally excluded through a combination of the
LEP Higgs and chargino search \cite{Barate:2003sz} and the requirement
of a neutral LSP\!. On the other hand, in the $\not\!\Psix$-SSM the
LSP need not be neutral and regions of no-scale mSUGRA parameter space
are allowed~\cite{Allanach:2003eb}.

%----------------------------------------------------------------------    

\section{The $\Psix$-Violating MSSM \label{sec:p6v}}

We next investigate the mass spectrum of the $\not
\!\!\!\Psix$-MSSM with particular focus on phenomenological analyses
in preparation for the LHC. The superpotential is given in
Eqs.~(\ref{superpot})-(\ref{P6v-superpot}) and has 48 additional
parameters, $\lam_{ijk},\,\lamp_{ijk},\,\lampp_{ijk},$ $\kap_i$, with
further parameters when we include SUSY breaking. Again, this
parameter space is too large for a detailed phenomenological analysis
and we restrict ourselves to the simplification of minimal
supergravity with universal soft SUSY breaking parameters at $M_X$.
The $\not\!\Psix$-mSUGRA model was developed in Refs.~\cite{Hempfling:1995wj,deCarlos:1996du,Allanach:1999mh,Allanach:2003eb}. The
RGEs connecting $M_X$ with the weak scale are given at one-loop in
Refs.~\cite{deCarlos:1996du,Allanach:1999mh,Besmer:2000rj,Allanach:2003eb}. 
The two-loop RGEs for the SM gauge couplings and the
superpotential parameters \cite{Allanach:1999mh} can have a
qualitative impact on gauge coupling and $b$-$\tau$ unification and are
included in our analysis. The impact of three-loop RGE terms is usually
1-2$\%$~\cite{jj2} for the squarks, and negligible for weakly
interacting sparticles.
The two-loop RGEs for the soft-breaking terms have recently been
calculated~\cite{jj}. 
We restrict ourselves to one-loop RGEs for the soft breaking parameters in
order to perform scans in parameter 
space in a reasonably small amount of CPU-time. This is likely to be an
excellent approximation for $\nPsixU$ couplings $\lsim 0.1$.

\mymed

In order to keep the parameter
space manageable, we shall also only consider one non-zero parameter
of $W_{\not\Psix}$ at a time at $M_X$. However, we include the
possible dynamical generation of the other parameters through the
coupled RGEs \cite{Allanach:2003eb}.  Note that at any given scale,
the $\kap_i$ can be rotated away through a redefinition of the fields
$L_i, H_1$ \cite{Hall, Dreiner:2003hw, Allanach:2003eb}.  Within the
context of supergravity with universal SUSY breaking we find it
natural to rotate the $\kap_i$ to $0$ at $M_X$ \cite{Allanach:2003eb}.
Non-zero $\kap_i$ are then dynamically generated through the RGEs,
leading to naturally small neutrino masses \cite{Nardi:1996iy,
Allanach:1999mh, Allanach:2003eb,Hempfling:1995wj} in agreement with
observations \cite{Rimmer}. The parameters of $\not\!\!\Psix
$-mSUGRA are thus given by
\bea 
&&\mzero\,,\, \mhalf\,,\,
\azero\,,\,\tanb\,,\, \sgnmu\,,\,\BLam \quad @ M_X,
\label{P6V-param} \\[1.7mm]
&&\mathrm{where} \; \; \BLam\in\{\lam_{ijk},\lamp_{ijk},
\lampp_{ijk}\}\,, \notag
\eea
and we emphasise that in any given model only \textit{one} non-zero
parameter for $\BLam$ is chosen at $M_X$. In the no-scale mSUGRA case,
although $\mzero=\azero=0$ at $M_{GUT}$, they are non-zero at the weak
scale through the RGE running.

\mymed

Given the additional couplings in $W_{\not\Psix}$ as well as the
corresponding soft SUSY breaking terms, we have the following changes
in the phenomenology compared to $\Psix$-mSUGRA.
\begin{enumerate}
\item Through the additional ${\not\!\Psix}$ couplings the RGEs change
and thus the spectrum and couplings change at
low-energy~\cite{Allanach:2003eb,jj}. 

\item For $\not\!\Psix$ the LSP is unstable and the cosmological bound 
\cite{Ellis:1983ew} no longer applies. Thus any supersymmetric particle 
can be the LSP
\beq 
\mathrm{LSP}\;\in \;\{\tilde\nu_L,\,\tilde\ell_{L,R}^\pm,\,\tilde
q_{L,R},\,\tilde\chi^0_1,\,\tilde\chi^\pm_1,\,\tilde g \}\,.
\label{lsp}
\eeq
Within the $\not\!\!\Psix$-mSUGRA model the nature of the LSP is
determined through the initial conditions at $M_X$,
Eq.~(\ref{P6V-param}), and the RGEs.

\item Through the $W_{\not\Psix}$-couplings, sparticles
  can be singly produced at colliders \cite{single-prod},
  \textit{e.g.} resonant slepton production at hadron colliders
  \cite{res-slep}.

\item Due to the possible changes in the spectrum as well as the
  additional interactions, the decay patterns of the supersymmetric
  particles can change, see for example Ref.~\cite{decay-patterns}.

\end{enumerate}

It is the purpose of this paper to investigate points 1 and 2 in
detail. Points 3 and 4 will be exemplified in benchmark points later,
but their detailed study will be left for later work. First we review
the experimental bounds on the couplings $\BLam$, and consider
experimentally excluded regions of parameter space.

\subsection{Bounds on the $\BLam$ Couplings \label{bounds}}

The low-energy 2$\sigma$ bounds on the tri-linear $\nPsixU$-couplings are
summarised in Refs.~\cite{barger,Dreiner:1997uz,allanach-dedes}. In
Table~\ref{ew-bounds} we show the largest allowed coupling in the five
``classes'' of couplings for scalar masses of 100 GeV. We do not include 
bounds which depend strongly on assumptions about quark mixing, such as those
from 
RGE-induced neutrino masses, but rather those which come from more direct
sources listed in Ref.~\cite{allanach-dedes}. 
\begin{table}
\begin{tabular}{|ccccc|} \hline
$\;\;\lam_{ijk}\;$ & $\;\lamp_{1jk}\;$ & $\;\lamp_{2jk}\;$ & 
$\;\lamp_{3jk}\;$ & $\;\lampp_{ijk}$  
\\[1mm] \hline
\;0.07&0.28&0.56&0.52& pert.\; \\[1mm] \hline
\end{tabular}
\caption{Weakest 2$\sigma$ direct bound on the five classes of couplings from 
  low-energy processes such as a measurement of the Fermi constant 
  in muon decay. Most of the bounds scale with $(\tilde m$/100$)\gev$, where
  $\tilde m$ is the relevant scalar fermion mass for the process
  leading to the bound. ``pert.'' indicates that the weakest bound only comes
  from perturbativity of the coupling $\mathcal{O} (1)$. Only one non-zero
  $\not\!\Psix$   coupling at a time is considered.}
\label{ew-bounds} 
\end{table}
The bounds shown are to be applied at $M_Z$, and no assumption about
the high energy completion of the model has been made. We see that the
weakest $\lampp_{ijk}$ bounds are not constraining with only
perturbativity as a limiting effect ($<$3.5 by some definitions).  It
should be noted that three of the $\lampp$ couplings involving the
first two generations have quite severe bounds~\cite{allanach-dedes}.
The weakest $\lamp_{ijk}$ bounds also appear to allow large $\sim
{\mathcal O}(1)$ couplings, but in fact many of the bounds are at the
10$^{-1}$ level. Finally, the $\lam_{ijk}$ bounds are all around $5
\cdot10^{-2}$.

\mymed

However, within the context of a supergravity model formulated at the
unification scale, we must take into account the $\not\!\Psix$-RGEs.
Considering the effect of the gauge couplings on the running of an
individual $\BLam$ strengthens the bounds by a factor $\sim(1.5, 3.5,
4.3)$ at $M_X$ for the couplings $(\lam,\lamp,\lampp)$, respectively
\cite{deCarlos:1996du,allanach-dedes}. The RGEs for the $\BLam$'s are
highly coupled. In particular, the phenomenological requirement of
$\mu\not=0$ together with some $\lam$ or $\lamp\not=0$ at $M_X$ leads
to non-zero $\kap_i$ at the electroweak scale \cite{Nardi:1996iy,
Allanach:1999mh, Allanach:2003eb}. Through mixing of the neutrinos
with the neutralinos, $\kap_i\not=0$ implies a non-zero neutrino mass
\cite{Hall}. Applying the cosmological bound on the neutrino mass, 
the resulting upper bounds on the $\lam_{ijk}(M_X),\,\lamp_{ijk}(M_X)
$ were obtained in Ref.~\cite{Allanach:2003eb}. The corresponding
weakest bounds at $M_X$ on the four classes of couplings are shown in
Table~\ref{neut-bounds} for the SPS1a point. Here, we have updated the
bounds using the latest {\tt SOFTSUSY} version \cite{Allanach:2001kg}.
Where the neutrino mass bound is not constraining, there is an upper
bound coming from the non-generation of negative mass squared scalars
via RGE effects. Such ``tachyon'' bounds are denoted by a superscript
``$x$'' in the table. Most of the bounds strongly depend on the origin
of CKM mixing in the SM quark sector \cite{CKM,allanach-dedes}. We see
from the table that if the CKM mixing originates solely in the
down-quark sector we obtain much stricter bounds than solely up-quark
sector mixing.
\begin{table}
\begin{tabular}{|c|cccc|} \hline
&$\lam_{ijk}\;$ & $\lamp_{1jk}\;$ & $\lamp_{2jk}\;$ & 
$\lamp_{3jk}\;$ 
\\[0.9mm] \hline
$\;d$-mixing\; & \;0.046\; & $\;9.1\cdot10^{-4}\;$ & $\;9.1\cdot10^{-4}\;$ & 
$\;9.0\cdot10^{-4}\;$ \\[0.9mm]\hline
$u$-mixing &0.046& $0.15^x$&$0.15^x$&$0.15^x$ \\[0.9mm] \hline
\end{tabular}
\caption{Weakest bound at $M_X$ on the four classes of lepton number 
violating couplings from the RGE generated $\kap_i$ for SPS1a. The
superscript $x$ denotes cases where there is no neutrino mass bound on
some couplings in that class and so we have indicated the bound coming
from the lack of tachyons. In the case of $\lam$ couplings the weakest
bound arising from the RGEs is $\lam_{231}<0.55^x$ from the lack of
tachyons \cite{allanach-dedes}. We have instead included the
translation of the weak scale bound to $M_X$. The weak scale bound
scales with the slepton mass $(m_{\tilde\ell}/100\,\mathrm{GeV})$ and
is stronger than the tachyon bounds for slepton masses below the TeV scale.}
\label{neut-bounds}
\end{table}
The nature of the bounds within the supergravity context is important
for models of flavour formulated at the GUT scale
\cite{Dreiner:2003hw}.

\section{Bounds on $\mathbf{\nPsixU}$ No-Scale 
Supergravity Parameter Space}
\label{sec:space}

Having discussed the bounds on the $\BLam$ couplings we next
consider the constraints on $\nPsixU$-mSUGRA parameter space
arising from precision measurements. Specifically, we consider the
anomalous magnetic moment of the muon $(g-2)$, the branching ratio
$\Br(b\ra s\gamma)$, and the branching ratio $\Br(B_s
\ra\mu^+\mu^-)$. We also briefly discuss the impact of the LEP Higgs
mass bound.

\mymed

All of the numerical calculations in the present paper have been
performed using an unpublished $\s \Psix$ version of \texttt{SOFTSUSY}
\cite{Allanach:2001kg}. The predictions for the $\Psix$ conserving
$\text{Br}_{\Psix}(B_s\ra\mu^+\mu^-)$, $(g-2)_\mu$, and $\text{Br}_{\Psix}(b\ra s\gamma)$ are calculated using
\texttt{micrOMEGAs1.3.6} \cite{micromegas}.  Throughout this paper, we
use the default inputs $M_Z=91.1875$ GeV and $m_t=172.5$ GeV for the
pole masses of the $Z^0$ boson and top quark, respectively. The
weak-scale gauge couplings are set to their central values in the
$\overline{MS}$ scheme $\alpha^{-1}(M_Z)=127.918$,
$\alpha_s(M_Z)=0.1187$. Light quark masses are also set to their
central values in the $\overline{MS}$ scheme: $m_b(m_b)=4.25$ GeV,
$m_u(2 \mbox{~GeV})=0.003$ GeV, $m_d(2 \mbox{~GeV})=0.00675$ GeV,
$m_s(2 \mbox{~GeV})=0.1175$ GeV and $m_c(m_c)=1.2$
GeV~\cite{Eidelman:2004wy}.

\mymed

\begin{figure}[t!]
\begin{center} 
\unitlength=1in 
\begin{picture}(3,2.3)
\put(-0.6,0){\epsfig{file=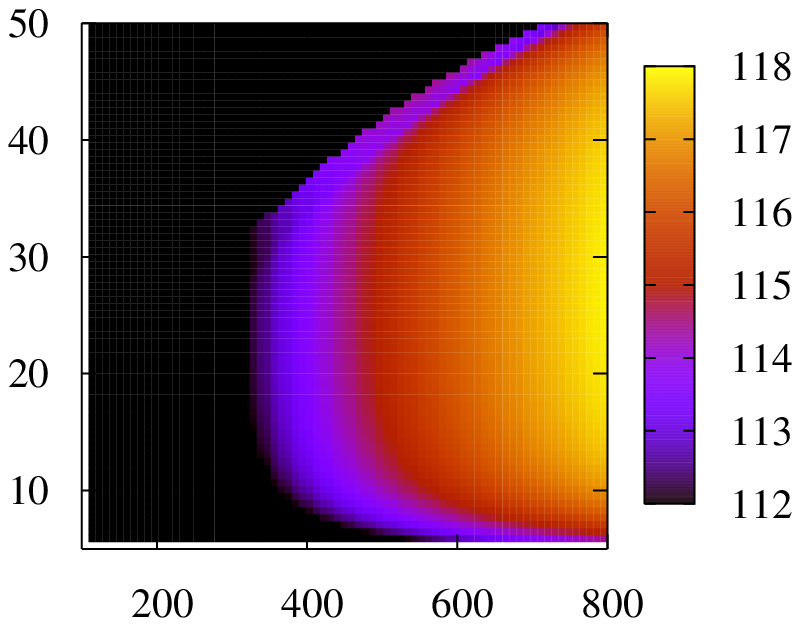, width=3.7in}}
\put(2.7,1.){\rotatebox{90}{$m_{\higgs}$ [GeV]}}
\put(1.3,0.2){\makebox(0,0){$\mhalf$ [GeV]}}
\put(0.05,1.2){\rotatebox{90}{$\tanb$}} 
\end{picture} 
\end{center}
\caption{Mass of $h^0$ in no-scale mSUGRA parameter space. The blacked
out region corresponds to points that fail negative LEP2 Higgs
constraints or that possess tachyonic scalars.}  \label{fig:higgsmass}
\end{figure}

The prediction for the lightest CP-even Higgs mass, $m_h$, is shown in
Fig.~\ref{fig:higgsmass} as a function of the no-scale mSUGRA
parameter space. As discussed above, a lower bound of 111.4 GeV is
imposed upon {\tt SOFTSUSY}'s prediction of $m_h$. We find $m_h<124
$ GeV for values of $M_{1/2}<1500$ GeV.

\subsection{THE MUON $\mathbf{(g-2)}$ VALUE}

The anomalous magnetic moment of positively and negatively charged
muons is one of the very few predictions of the SM that are
\textit{not} in good accordance with experiment. The combined
experimental measurement is~\cite{Bennet} \beq a_\mu^{\text{exp}}
\equiv \frac{(g-2)_\mu}{2} =11659208.0(6.3) \times 10^{-10}. \eeq The
SM prediction requires the input of data on hadronic spectral
functions.  There is a well-known discrepancy between the predictions
based on data from $e^+e^-\!\rightarrow\,$hadrons (``$e^+e^-$-based'')
and those based on semi-leptonic tau decays (``tau-based'')
\cite{davier}. Following the particle data group \cite{marciano}, we
focus on the $e^+e^-$-based predictions and obtain
\beq
\delta a_\mu \equiv a_\mu^{\mbox{exp}} - a_\mu^{\mbox{SM}}
= (22.2\pm 10.2) \times 10^{-10}, \label{gm2}
\eeq
where all theoretical and experimental errors have been added in
quadrature. $\delta a_\mu$ is smaller and within one sigma from zero
when the $\tau$-based prediction is used. We shall constrain the
$\nPsixU$-MSSM such that the predicted $\delta a_\mu^{\text{SUSY}}
\equiv a_\mu^{\text{exp}}-a_\mu ^{\text{SUSY-th}}$ provides the
missing component of $(g-2)_\mu$ in Eq.~(\ref{gm2}).

\begin{figure}[t!]  
\begin{center} 
\unitlength=1in
\begin{picture}(3,2.3) 
\put(-0.6,0){\epsfig{file=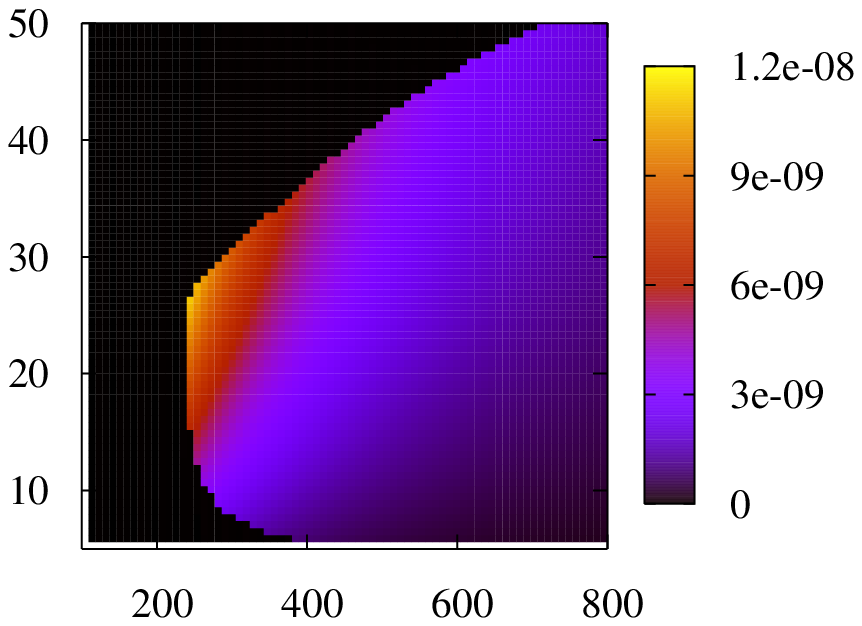,width=3.7in}} 
\put(0.33,0.48){\epsfig{file=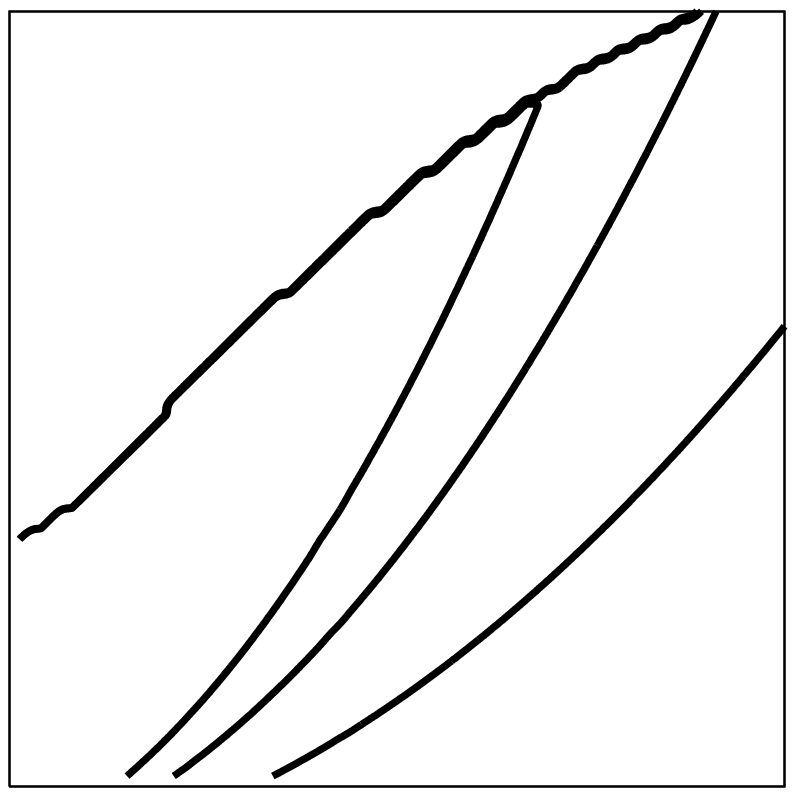,width=2.509in}} 
\put(2.9,1.3){\rotatebox{90}{$\delta a_{\mu}$}}
\put(1.3,0.2){\makebox(0,0){$\mhalf$ [GeV]}}
\put(0.05,1.2){\rotatebox{90}{$\tanb$}} 
\end{picture} 
\end{center}
\caption{No-scale $\Psix$ contribution to $\delta
  a_{\mu}^\text{SUSY-th}$, \textit{i.e.} $\mzero=\azero=0$ at $M_X$.
  The blacked out region to the left and upper left is excluded due to
  LEP exclusion bounds and tachyons.  The black contours show the
  region of parameter space that is consistent with the empirical
  central value of $g-2$, plus or minus 1$\sigma$.}
\label{fig:gmum12tanb}
\end{figure}

\mymed

In order to compute $\delta a_\mu^\text{SUSY-th}$, we use {\tt
micrOMEGAs1.3.6} \cite{micromegas} for the one-loop supersymmetric
$\Psix$ part, which can be enhanced by $\tan\beta$. The one-loop
$\nPsixU$ contribution is of order 
\beq
(\delta a_\mu)^{\nPsixU} \sim \mathcal{O} \left(
\frac{m_\mu^2 {|\BLam|}^2}{32 \pi^2 {\tilde m}^2}\right),
\eeq
for $\BLam\in\{\lam_{ijk},\lamp_{ijk}\}$~\cite{kimetal}, where $\tilde
m$ is a sparticle mass that appears in the loop. For $\tilde m>300$
GeV and $|\BLam|^2 < 0.1$ (as is the case for the points studied
here), $(\delta a_\mu)^{\nPsixU}<\mathcal{O}(10^{-1 1})$ and so we
neglect it, taking the $\Psix$ contribution only.

\mymed

Fig.~(\ref{fig:gmum12tanb}) shows the allowed region of $\delta
a_{\mu}^{\text{SUSY-th}}$ in the no-scale MSSM in the $\mhalf$-$\tanb
$-plane as the region between the contours. At higher SUSY masses
(higher $\mhalf$), $\delta a_{\mu}^\text {SUSY-th}$ becomes smaller as
the SUSY contributions decouple due to mass suppression of sparticle
propagators in the one-loop diagrams.  Here, the prediction of $a_\mu$
is similar to that of the SM. $\delta a_\mu$ increases for higher
$\tan\beta$, which is why the contours move to the right with
increasing $\tan\beta$.

\subsection{\bf{$\Br(b\ra s\gamma)$}}
    
\begin{figure}[t!]
      \begin{center} \unitlength=1in \begin{picture}(3,2.3)
      \put(-0.6,0){\epsfig{file=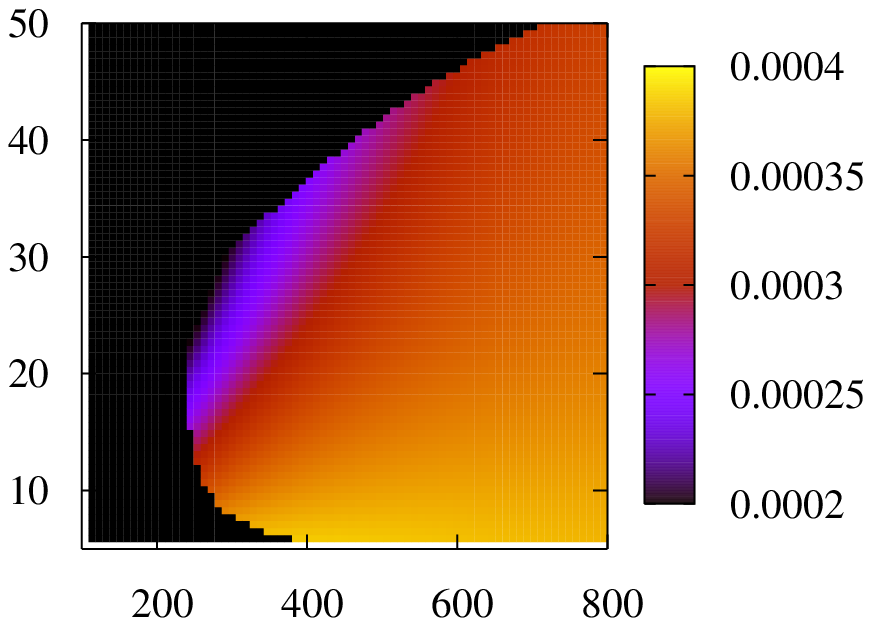, width=3.7in}}
      \put(0.33,0.48){\epsfig{file=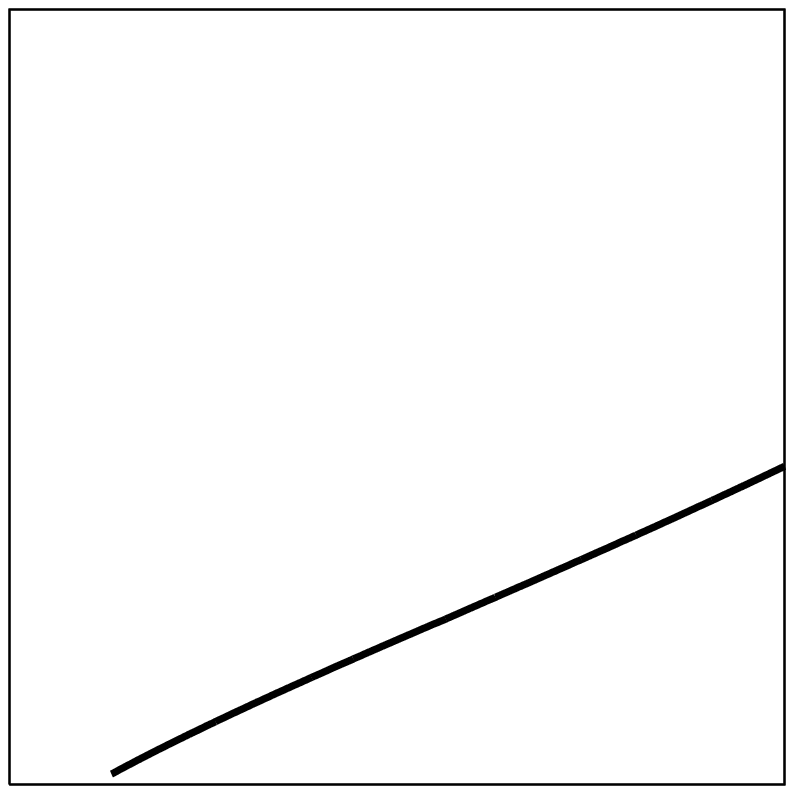, width=2.509in}}
      \put(2.9,1.1){\rotatebox{90}{${\mathrm Br}(b\ra s\gamma)$}}
      \put(1.3,0.2){\makebox(0,0){$\mhalf$ [GeV]}}
      \put(0.05,1.2){\rotatebox{90}{$\tanb$}} 
\end{picture}
\end{center} 
\caption{$\Psix$-MSSM no-scale mSUGRA contribution to ${\mathrm {Br}}(b
      \ra s \gamma)$. The blackened out region is excluded due to LEP
      measurements and the existence of tachyons. The contour shows
      the mid-value of ${\mathrm {Br}}(b \ra s \gamma)$.}
      \label{fig:bsgammam12tanb}
\end{figure}

Next we consider the measurement of $\Br(b\ra s\gamma)$. The
current experimental result is~\cite{twentythree}
\beq
\Br(b\ra s\gamma)_{\mathrm{exp}}=(3.55\pm 0.26)\times 10^{-4},
\eeq
where we have added the statistical and systematic errors in
quadrature, including the uncertainty for the shape function of the
photon energy spectrum. If we now include a combined theoretical
error of 0.30$\times 10^{-4}$ \cite{twentyfour} and add it in
quadrature to the experimental error we obtain the allowed range for
the central theoretical value at the 2$\sigma$ level~\cite{allanach}
\beq
2.76\times 10^{-4} <\Br(b\ra s\gamma)_{\mathrm{th}}<4.34\times 10^{-4}.
\eeq
The $\Psix$-MSSM branching ratio is plotted in
Fig.~(\ref{fig:bsgammam12tanb}) for the no-scale mSUGRA model. Again,
$\mhalf$ and $\tanb$ are varied. The contour shows the mid-value of
$\Br(b\ra s\gamma)$, \textit{i.e.} $3.55\times
10^{-4}$.

\mymed
    
The $\Psix$ conserving contribution to $\Br(b\ra s\gamma)$ of
the no-scale mSUGRA model is thus in agreement with the experimental
bounds at $2 \sigma$ in almost the entire plotted region. An estimate
for the $\nPsixU$ contribution to this process is given in
Ref.~\cite{Besmer:2000rj}. There are many potential contributions from
products of two different $\BLam$ couplings. Although we assume only
one non-zero $\BLam$ coupling at the GUT scale, several different
couplings are generated by the RGE evolution to the weak scale
providing a non-zero $\nPsixU$ contribution. However, the couplings
generated by the RGEs are suppressed by loop factors such that the
$\nPsixU$ generated contribution to $\Br(b\ra s \gamma)$ is
also highly suppressed. We have generalised Ref.~\cite{Besmer:2000rj}
to include CKM quark mixing and explicitly checked that the
contribution to $\Br(b \ra s \gamma)$ is less than $10^{-6}$
in every case studied and therefore negligible.  For example, the
difference between the predicted values of $\Br(b \ra s
\gamma)$ in the $\Psix$ case and the combined $\Psix$ and $\nPsixU$
case for the no-scale mSUGRA point $M_{1/2}=400$ GeV, $\tan \beta=20$
is always smaller in magnitude than $10^{-9}$ for quark mixing purely
in the down sector and for any $\lamp_{ijk}$ set at its upper limit. A
similar conclusion holds in the case of up-mixing.

\subsection{\bf{$\Br(B_s\ra \mu^+\mu^-)$}}

The Tevatron experiments have recently tightened the upper limit on
the rare decay branching ratio $\Br(B_s\ra \mu^+\mu^-)$,
currently the best CDF bound is~\cite{tevbsmumu}
\beq
\Br_{\mathrm{exp}}(B_s\ra \mu^+\mu^-)<1\times 10^{-7}
\label{cdf-bsmumu}
\eeq
at the 95$\%$ C.L. 
The SM prediction is~\cite{buras}: 
\beq
\Br_{SM}(B_s\ra
\mu^+\mu^-)=(3.35 \pm 0.32)\times 10^{-10}\,.
\eeq

\begin{figure}[t!]
\begin{center}
\unitlength=1in
\begin{picture}(3,2.3)
	  \put(-0.6,0){\epsfig{file=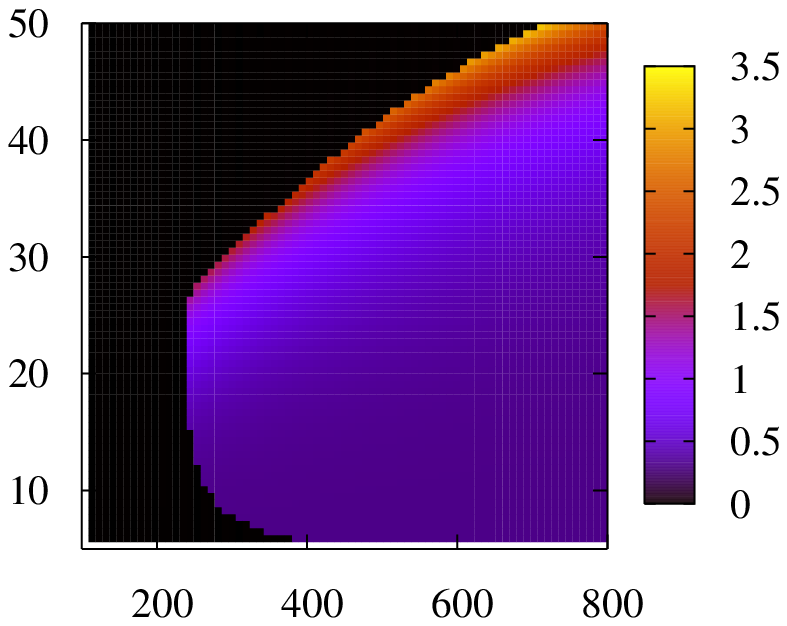, width=3.7in}}
  \put(2.7,0.65){\rotatebox{90}{${\Br}_{RPC}(B_s \ra \mu^+\mu^-)
      /10^{-8}$}}
	  \put(1.3,0.2){\makebox(0,0){$\mhalf$ [GeV]}}
	  \put(0.05,1.2){\rotatebox{90}{$\tanb$}}
	\end{picture}
      \end{center}
      \caption{$\Psix$ contribution to $\Br_{SM}(B_s\ra
        \mu^+\mu^-)$ in no-scale mSUGRA. The blacked out region is excluded by
        LEP direct sparticle and Higgs searches, as well as the requirement of
        no tachyons.
	\label{fig:bsmumu}}
\end{figure}
In the $\Psix$-MSSM, the branching ratio can be significantly larger
than in the SM since it is proportional to $(\tan\beta)^6/M_A^4$
\cite{choudhury,bobeth}. The $\tan \beta$ enhancement comes from the
dependence on bottom and muon Yukawa couplings. The above experimental
bound, Eq.(\ref{cdf-bsmumu}), constrains some of the $\mzero\neq 0$
mSUGRA parameter space~\cite{Dedes:2001fv}. In Fig.~(\ref{fig:bsmumu})
we show the $\Psix$-mSUGRA prediction for $\Br(B_s\ra\mu^+\mu
^-)$ for the specific parameter range of the \textit{no-scale} mSUGRA
model, \textit{i.e.} with $\mzero=0$, as a function of $ \mhalf$ and
$\tan\beta$.  We see that in this case, it is at least an order of
magnitude below the experimental bound over the whole plane.
    
\mymed

Since the $\Psix$-mSUGRA contribution is small, we look at the
$\nPsixU$ contribution alone, \textit{i.e.} without
interference with the SM or $\Psix$ conserving contribution. The $
\nPsixU$ contribution to $\Br(B_s\ra\mu^+\mu^-)$ was calculated in
Ref.~\cite{Jang:1997jy}. Here, we generalise the calculation in order
to include the L-R--mixings of the mediating squarks and also allow
for non-degenerate sparticle masses. The computation is presented in
the Appendix. We use this generalised result in an unpublished
$\nPsixU$ version of \texttt{SOFTSUSY} to give an estimate of the
contribution to the decay rate. In the case of one dominant $\nPsixU$
operator at the GUT scale, the contribution to $\Br(B_s\ra
\mu^+\mu^-)$ is always less than $10^{-14}$ once the $\BLam$ have been
bounded by experimental constraints, irrespective of the nature of the
CKM mixing. For our analysis, the $\nPsixU$ contribution to
$\Br(B_s\ra\mu ^+\mu^-)$ is therefore neglected in the rest of the
paper. However, we note that if we drop the requirement of only one
non-zero GUT-scale $\BLam$ coupling, much larger $\nPsixU$
contributions are possible. For example, choosing $\lam_{122}(M_{GUT}
)=6.3 \times 10^{-4}$ and $\lamp_{132}(M_{GUT})=7.6\times 10^{-5}$
achieves a $\nPsixU$ contribution to $\Br(B_s\ra \mu^+\mu^-)$ as high
as $10^{-7}$.  The corresponding product bounds for weak-scale
couplings are given in Ref.~\cite{Jang:1997jy}

\section{$\not\!\Psix$-Spectrum \label{sec:spec}}
In this section, we investigate the supersymmetric spectrum in the
$\not\!\Psix$-mSUGRA  model, taking into account the
constraints of the previous section. We are particularly interested in
the nature of the LSP, which is of fundamental importance for the
resulting collider signatures. However, we first investigate the
overall shift in the supersymmetric spectra due to the $\BLam$
couplings. Finally, we study the changes in the mass ordering of
sparticles. The latter determines the cascade decay patterns, which in
turn are also essential for the collider signatures.

\subsection{Shift in Spectrum due to $\BLam$-RGEs}

In order to determine the low-energy $\not\!\!\Psix$ supersymmetric
spectrum from the parameter set at $M_X$,
Eq.~(\ref{P6V-param}), we must employ the $\not\!\!\Psix$ RGEs
\cite{deCarlos:1996du,Allanach:1999mh, Besmer:2000rj,
  Allanach:2003eb}. They have been programmed at one-loop order in an
unpublished version of \texttt{SOFTSUSY} \cite{Allanach:2001kg}. The
leading effect of the $\BLam$ couplings on the running of the
supersymmetric masses is $\sim\mathcal{O}[\BLam^2/( 16 \pi^2)\ln(M_{GU
  T}/M_Z)]$. Given the bounds in Table~\ref{neut-bounds}, in most
cases we only expect small shifts in the spectrum in the mSUGRA
scenario. In order to illustrate the effects, we choose as a
comparison the low-energy spectrum of the SPS1a point
\cite{Allanach:2002nj} with conserved $\Psix$. It is listed in the
second column of Table~\ref{spectra} and also shown in
Fig.~\ref{fig:sps1a}. In the first column we have listed the
supersymmetric particles.

\begin{table}
  \begin{tabular}{|c|c|c|c|c|} \hline
    & $\Psix$ & $\lam_{123}= 0.08$ & $\lamp_{331}=0.122 $ & $\lampp_{212}=0.5$
    \\ \hline
    $\;\tilde\nu_e\;$     & 189     & 187 & 189        & 189\\ \hline
    $\;\;\tilde\nu_\mu\;\;$   & 189     & 187 & 189        & 189\\ \hline
    $\;\;\tilde\nu_\tau\;\;$  & 188     & 188 & \textbf{93}  &  188\\ \hline
    $\tilde e^\pm_{R,L}$  & 146; 206 & 146; 205 & 146; 206& 146; 206\\
    \hline
    $\;\tilde \mu^\pm_{R,L}\;$ & 146; 206 & 146; 205 & 146; 206 & 146; 206\\
    \hline
$\tilde\tau^\pm_{1,2}$&137; 210&134; 210&\textbf{104}; \textbf{159}&137; 210\\ \hline
    $\tilde u_1 $         & 552      & 552     & 552      & 552  \\ \hline
    $\tilde u_2 $         & 567      & 567     & 567      & 568  \\ \hline
    $\tilde c_1 $         & 552      & 552     & 552      & \textbf{394}\\\hline
    $\tilde c_2 $         & 567      & 567     & 567      & \textbf{562} \\ \hline
    $\tilde d_1$          & 552     & 552    & \textbf{536} & \textbf{393}\\\hline
    $\tilde d_2 $          & 575     & 575     & 574      & \textbf{570} \\ \hline
    $\tilde s_1 $         & 552      & 552     & 552 & \textbf{393}\\ \hline
    $\tilde s_2 $         & 575      & 575     & 575      & \textbf{570} \\ \hline
    $\tilde b_{1,2}$    & \;518; 550\; & 518; 550& \textbf{511}; 549 & 519; 551 \\
    \hline
    $\tilde t_{1,2}$      & 400; 591  & 400; 591& 399; \textbf{586} & 401; 592 \\
    \hline
    $\tilde\chi_1^0$            & 97       & 97       & 97       & 97 \\ \hline
    $\tilde\chi_2^0$            & 181      & 181      & 181      & 181 \\ \hline
    $\tilde\chi_3^0$            & 362      & 362      & 360      & 362 \\ \hline
    $\tilde\chi_4^0$            & 380      & 380      & 379      & 380 \\ \hline
    $\tilde\chi_1^\pm$          & 182      & 182      & 181      & 182 \\ \hline
    $\;\tilde\chi_2^\pm\;$      & 378      & 378      & 377      & 378 \\ \hline
    $\tilde g$            & 610      & 610      & 610      & \textbf{604}
    \\ \hline
    $h^0$                 &  110     & 110      & 110      & 110 \\ \hline
    $H^0$                 & 397      & 397      & 396      & 397 \\ \hline
    $H^\pm$               & 405      & 405      & 404      & 405 \\ \hline
    $A^0$                 & 397      & 397      & 396      & 397 \\ \hline
\end{tabular}
\caption{Low-energy supersymmetric spectra for the point SPS1a in the
  $\Psix$ conserving case, column two, and in various maximum $\not\!
  \Psix$ violating scenarios in columns 3-5. In the latter other than 
  $\BLam(M_X)$ all SUSY parameters are as in SPS1a. The masses are 
  given in GeV and we have high-lighted in bold face those changes 
  compared to the $\Psix$-spectrum which are 5 GeV or more. In the case 
  of $\lamp_{331}$, we assume the entire CKM mixing is in the up-sector. 
  In the case of $\lam''_{212}$ we have assumed the CKM mixing is in the 
  down-sector.}
\label{spectra}
\end{table}

\mymed 

In the remaining columns, we show various $\not\!\!\Psix$ 
spectra resulting from one non-zero coupling at $M_X$ and with the
$\Psix$ conserving parameters as for SPS1a. In the third column of
Table~\ref{spectra}, we list the masses for the largest allowed $LL
\bar E$ coupling: $\lam_{123}(M_X)=0.08$. We see that the maximal
relative shift of any supersymmetric particle is about $2\%$ and is in
the $\tilde e_R$ mass.  For most particles the shift is below 1\%,
which is indistinguishable from the $\Psix$ spectrum at the level of
significance we display. In the fourth column, we list the spectrum
for $\lam^\prime_{331}(M_X)= 0.122$, \textit{i.e.}  the case of
up-mixing \cite{ftnt5}. Here we obtain significant shifts in the
masses of $\tilde\nu_\tau$ and $\tilde\tau_ {1,2}$, which directly
couple to the operator $\lam_{331}L_3Q_3\bar D _1$. In particular,
this leads to the novel possibility of a tau sneutrino LSP, which we
discuss in more detail below. There are also slight shifts in the
$\tilde d_1,\,\tilde b_1,\,\tilde t_2$ masses, with the operators
again coupling directly to the dominant $\not\!  \Psix$ operator. In
the case of down-mixing, $\lamp_{331}$ is bounded to be less than 
$10^{-3}$ and the spectrum is 
indistinguishable from the $\Psix$ spectrum in the second column and
we do not list it.

\mymed

The $\lampp$ couplings are not bounded by neutrino masses and in some
cases can still be at the perturbativity limit. As an example, we
choose $\lampp_{212}(M_X)=0.5$. We show the spectrum in the last
column on the right. Here we obtain significant shifts ($\sim 30\%$)
in the masses of $\tilde c_1,\,\tilde d_1, \,\tilde s_1$, which are
all dominantly $SU(2)$ singlets and thus directly couple to the
leading $\not\!\!\Psix$ operator. There are also small shifts in the
corresponding doublet masses as well as the gluino mass (originating
from changes to the 1-loop threshold contributions from the squarks).
We expect similar effects for the other large $\lampp$ couplings,
\textit{i.e.}  the fields which directly couple should experience the
largest mass shifts.

\subsection{Nature of the LSP}

As pointed out in Eq.~(\ref{lsp}), in the $\nPsixU$-MSSM in principle
any supersymmetric particle can be the LSP\@. However, given the
restricted parameter space of Eq.~(\ref{P6V-param}) at $M_X$, it is a
computational question as to which LSPs are attained in a given model.
In the general mSUGRA case and for vanishing $\BLam$-couplings, we
have the well-known (but usually ignored) region shown in the lower
right of Fig.~\ref{fig:triangle}, where the lightest stau ($\tilde
\tau_1$) is the LSP. For $A_0=-100\gev $, $\tan\beta=10$, $\sgnmu=+1$
the black line denoting the border between the $\tilde\chi_1^0$-LSP region,
at high values of $M_0$ and the $\tilde\tau_1$-LSP region at low
values, is approximately given by
\beq
M_{1/2}=3.8 \cdot M_0 + 175\gev\,.
\eeq
Such a $\,\tilde\tau_1$-LSP region always exists in mSUGRA models,
since for rising $M_{1/2}$ and fixed $M_0$ the gaugino masses grow
more quickly than the slepton masses, see the approximate formul\ae\
in Ref.~\cite{Ibanez:1983di}. However already for $\mzero\gsim200\gev
$, we must have $M_{1/2}\gsim1\,\mathrm{TeV}$ for a stau LSP. We can
read-off the no-scale case along the $x$-axis in
Fig.~\ref{fig:triangle}, where $\mzero=0$. This always results in a
$\tilde \tau _1$-LSP.

\mymed

If we now turn on the $\BLam$-couplings, this picture is in principle
modified~\cite{jj}. However as seen in the previous section, for the maximally
allowed couplings the shifts in the spectrum are typically modest for
all but the $\lampp$-couplings and exceptional cases of the $\lam'_
{331}$ coupling. For SPS1a, the charged sleptons are the next lightest
particles. Other than the stau, we thus might expect a $\tilde \mu_1$
LSP. However, we have checked that this requires $\lam_{ijk}$- or
$\lamp $-couplings far exceeding the upper bounds given in
Table~\ref{neut-bounds} \cite{ Allanach:2001kg}. (The $\lampp
$-couplings do not significantly affect the sleptons, see
Table~\ref{spectra}.)  In fact, in the case of down-mixing in the
weak-scale CKM matrix, we have found no further LSP. While the precise
numerical bounds on the couplings are sfermion mass dependent and were
only calculated for the SPS1a point \cite{Allanach:2003eb}, the
qualitative conclusion regarding possible identities of the LSP holds
in the region studied in Fig.~\ref{fig:triangle}.

\mymed 

\begin{figure}[t!]
\begin{center}
\unitlength=1in
\begin{picture}(3,2.3)
	  \put(0,2.12){{\epsfig{file=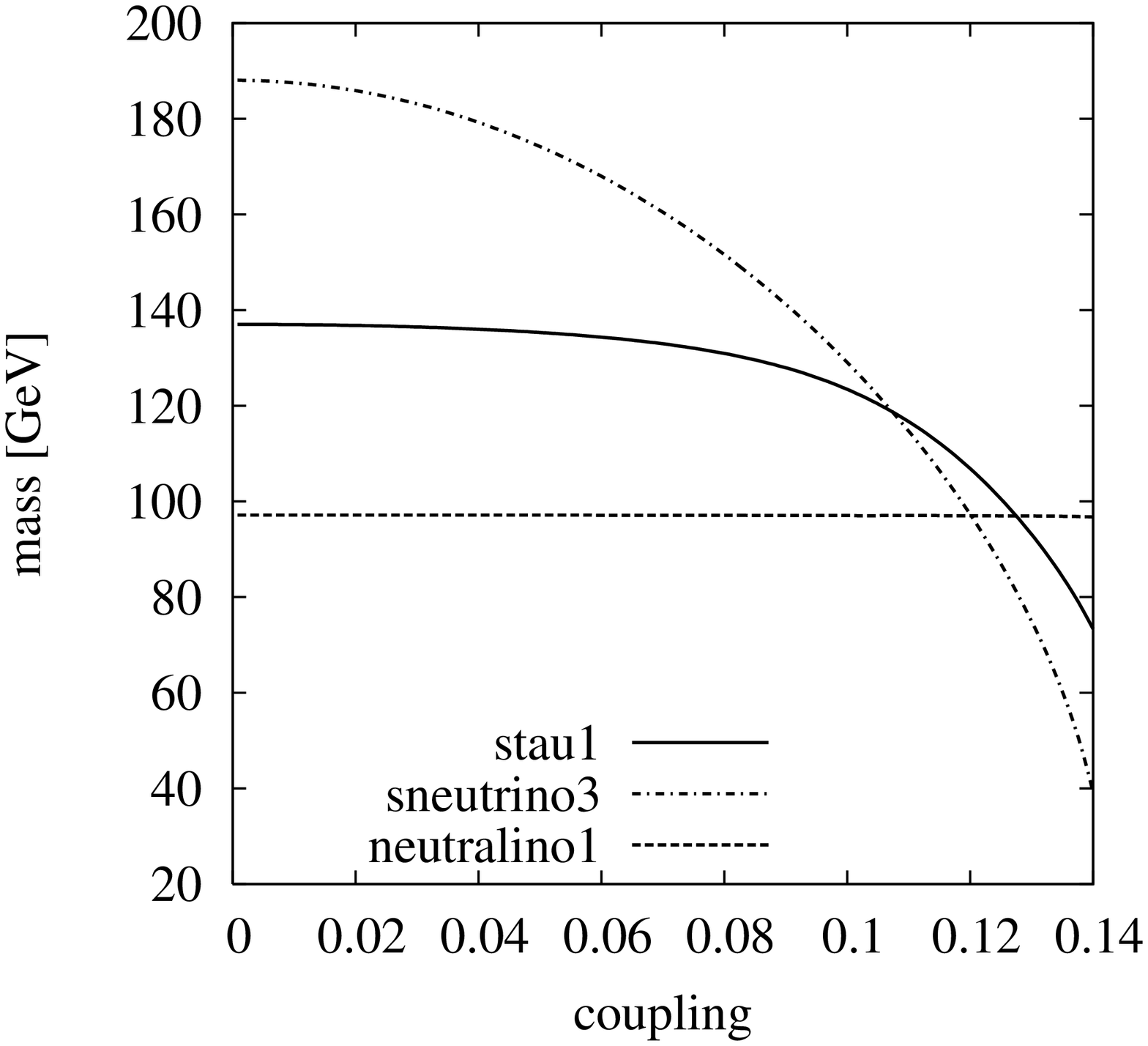, height=3.4in, angle=270}}}
	\end{picture}
      \end{center}
      \caption{LSP, NLSP and NNLSP at SPS1a for varying
      $\lamp_{331}(M_{GUT})$. All quark mixing has been placed in the up
      sector at the weak scale.
	\label{fig:snu}}
\end{figure}

The significant shifts in the spectrum in the $\lampp_{212}$ case
mainly affect the squarks, but they are too small to obtain a squark LSP.
However, for up-mixing in the case of non-zero $\lamp_{331}(M_{GUT})$,
it is possible to obtain a tau sneutrino LSP. As can be seen in
Fig.~\ref{fig:snu}, for the very narrow region $\lamp_{331}=0.12-0.14
$, a tau sneutrino LSP results at SPS1a. The lightest neutralino 
becomes the NNLSP for $\lamp_{331}>0.13$.

\mymed

To summarise, we have found that there is a well-motivated region in the
$\nPsixU$ no-scale supergravity model where the $\tilde\tau_1$ is the
LSP. This region has hitherto been almost completely ignored in the
literature \cite{Allanach:2003eb,stau}. We have also found a very
special region in parameter space where we have a tau sneutrino LSP,
\textit{i.e.} we have one of the central results of our paper
\beq
\mathrm{LSP}\in \left\{\,\tilde\chi^0_1,\,\tilde\tau_1,\tilde\nu\,\right\}\,.
\eeq
In the following, we investigate the spectrum in detail for the case
of a stau or sneutrino LSP. We then propose several $\nPsixU$
benchmark points for follow-up systematic phenomenological analyses,
including detector simulations.

\subsection{Mass Ordering in the Stau-LSP Region}

In the standard mSUGRA model, supersymmetric particles are produced in
pairs at colliders. They then cascade decay to the neutralino LSP
through various intermediate supersymmetric states. The LSP escapes
detection, which results in the typical missing transverse energy
signature for the $\Psix$-MSSM.

\mymed

We are here interested in the mSUGRA region with a $\stau_1$-LSP. If
the $\BLam$ couplings are small, as indicated by most of the bounds,
sparticles will be dominantly pair-produced in the usual $\Psix$
conserving channels \cite{ftnt2}. They will then cascade decay down to
the lighter supersymmetric particles eventually ending at the
$\tilde\tau_1$-LSP. The nature of these cascades and thus the final
state particles will be different from the $\Psix$-case since the
ordering of the spectrum can change. In particular, the lightest
neutralino will be heavier than one or more supersymmetric particle.
Finally, the resulting LSPs will decay into SM particles, the flavour
content of which depends on the dominant $\BLam$ coupling and again on
the sparticle mass ordering. We thus investigate in more detail
the mass ordering.

\mymed

\begin{figure}[t!]
\begin{center}
\unitlength=1in
	\begin{picture}(3,2.3)
	  \put(-0.6,0){\epsfig{file=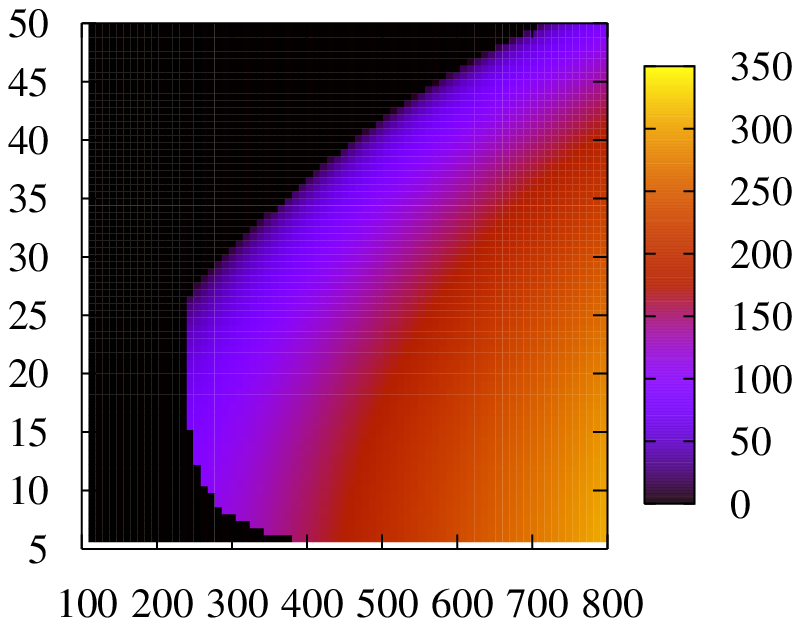, width=3.7in}}
	  \put(0.33,0.48){\epsfig{file=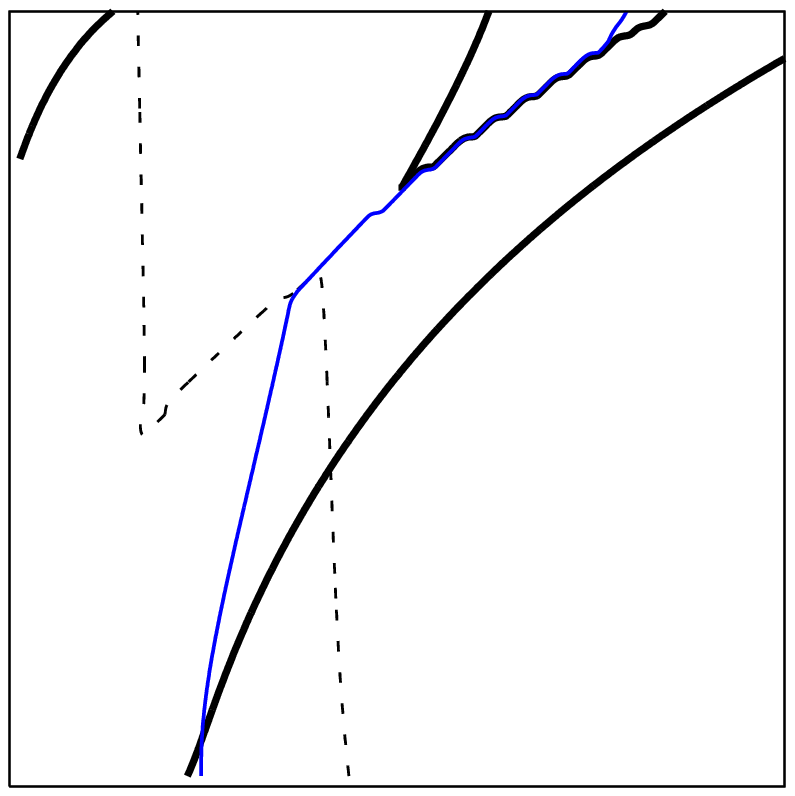, width=2.509in}}
	  \put(2.7,1.1){\rotatebox{90}{$m_{{\stau}_1}$ [GeV]}}
	  \put(1.3,0.2){\makebox(0,0){$\mhalf$ [GeV]}}
	  \put(0.05,1.2){\rotatebox{90}{$\tanb$}}
	  \end{picture}
\end{center}
\caption{Mass of the stau-LSP in the $\mhalf$-$\tanb$-plane for
  no-scale mSUGRA. The blackened out region is excluded due to LEP
  bounds and tachyons. The black contour shows $m_{h} = m_{\stau_1}$.
  The dashed contour shows $m_{\neutralino_1} =
  m_{\selectron_1}\approx m_{\tilde\mu_1}$, thus separating regions of
  $\neutralino_1$ NLSP (left) and $\neutralino_1$ NNNLSP (right). The
  blue contour shows where $m_{ \neutralino _2} = m_{\stau_2}$.}
\label{fig:staumzoom12tanb}
\end{figure}

Fig.~\ref{fig:staumzoom12tanb} shows the $\stau$-LSP mass in the
$\mhalf$-$\tanb$-plane for no-scale mSUGRA with $ \sgnmu =+1$.  We
shall mainly focus on the case of lepton-number violating couplings.
Here in the figure, the $\nPsixU$-couplings are set to zero, which
will be a good approximation for all cases where they are $\lsim$0.05.
The blackened out region to the left is excluded due to the LEP2 Higgs
search and the absence of tachyons. In the $\Psix$ limit, LEP2 gave an
86 GeV lower bound on the lightest stau mass \cite{Eidelman:2004wy},
which is also included. 

\mymed

The contours in the figure delineate regions differing in sparticle
mass ordering. The black contour shows where the mass of the lightest
CP-even Higgs, $m_{h^0}$, is equal to the mass of the $\stau_ 1$. For
models where $\mhalf$ is smaller, \textit{i.e.} to the left, $m_{h^0}
>m_{\tilde\tau_1}$. The dashed contour shows the analogous effect for
$m_{\neutralino_1}= m_{\tilde e_ 1}$, \textit{i.e.} to the right the
neutralino is heavier than the lightest selectron.  The neutralino is
then also heavier than the lightest smuon, since $m_{\tilde\mu_1}
\approx m_{\tilde e_1}$, with the former slightly lighter. We thus
have an extensive region, where $\tilde\chi^0_1$ is the fourth lightest
supersymmetric particle, the NNNLSP.  This should particularly have an
impact on slepton pair production signatures.  For squark pair production, the
heavier neutralinos and the two charginos which can appear in the
cascade decay can in principle have significant branching ratios,
where they completely skip the lightest neutralino directly decaying
to the light sleptons.

\mymed

It is also worth pointing out that for high $M_{1/2}$ and high $\tan
\beta$, we get the following additional changes compared to the SPS1a
spectrum.
\bea
m_{\neutralino_2} &>&  m_{\stau_2} \\
m_{\tilde \nu_3} &<& m_{\stau_2} < m_{\tilde\nu_{1,2}}\,.
\eea

\mymed 

In summary, we expect a significantly changed phenomenology in the
no-scale $\tilde\tau_1$-LSP parameter space, both compared to the
$\Psix$ case but also compared to the hitherto extensively studied
$\nPsixU$ case (so far usually called the R-parity violating case)
with a neutralino LSP. In order to investigate this in more detail in
the future, we propose several benchmarks which are representative of
the possible collider signatures.

\section{$\not\!\Psix$-BENCHMARKS \label{sec:bench}}

We now turn to the description of the $\nPsixU$-mSUGRA benchmarks. In
order to fully specify the model, we must fix both the standard mSUGRA
parameters: $M_{1/2},\,\mzero,\,\azero,\,\tan\beta,\,\sgnmu$, as well as
the $\BLam$ couplings. 
%%%%%%%%%%% BEN
%As the $\BLam$ couplings are somewhat independent
%from the other parameters, we
%first show three spectra, labelled I, II, and III, resulting from the
%former.
%%%%%%%%%%%
We first discuss the spectra of the benchmark points and then go on to detail
sparticle decays.

\subsection{Sparticle Spectra}

\begin{figure}[ht!]
      \scalebox{.8}{\include{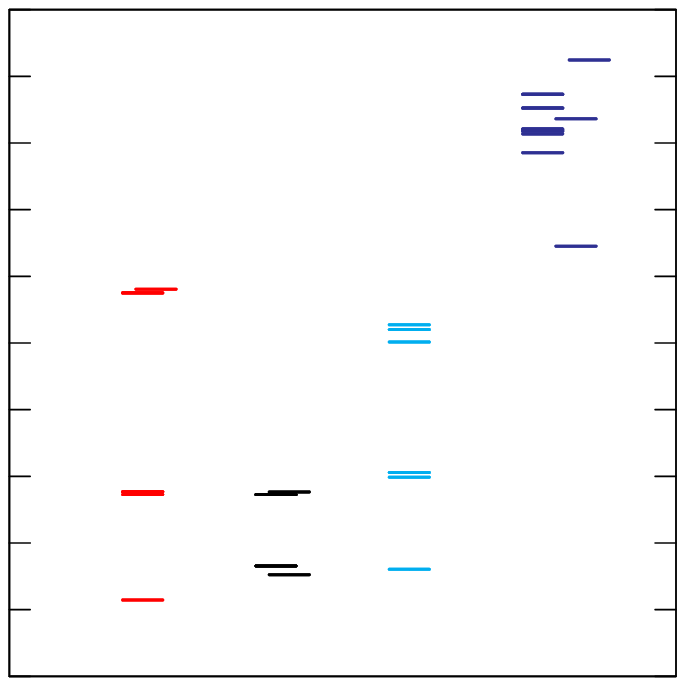}}
      \caption{Sparticle spectrum for no-scale mSUGRA parameter set:\;\;
      $\mhalf=400$ GeV, $\tanb=13$, $\sgnmu=+1$ and $\BLam =0$. \label{fig:benchI}} 
\end{figure}
The first 
no-scale mSUGRA parameter set consists of a rather light sparticle
spectrum: $M_{1/2}=400$ GeV, $\tan\beta=13$ and $\sgnmu=+1$. The
spectrum is shown in Fig.~\ref{fig:benchI}. As can be seen by a
comparison with Fig.~\ref{fig:sps1a}, the spectrum is similar to
SPS1a. The most important difference is a $\stau$-LSP of mass 148.2
GeV.  The selectron and smuon are the next lightest sparticles, with
masses 160.8 and 161.3 GeV respectively. The lightest neutralino is
the NNNLSP with a mass 161.5 GeV, almost degenerate with the lightest
charged sleptons, such that the direct decays are suppressed. The
sparticles are light so that there should be copious SUSY particle
production at the LHC. We have calculated this spectrum in the limit
of zero $\BLam$ and the spectrum in Fig.~\ref{fig:benchI} will be a good
approximation when all $\BLam \ll 1$, as is the case for two of our benchmarks to
be specified below: BC1 and BC2.

\mymed
%
%%%%%%%%%%%%%%%%%%%%%%%%%%%%%%%%%%%%%%%%%%%%%%%%%%%%%
\begin{figure}[ht!]
   \scalebox{.8}{\include{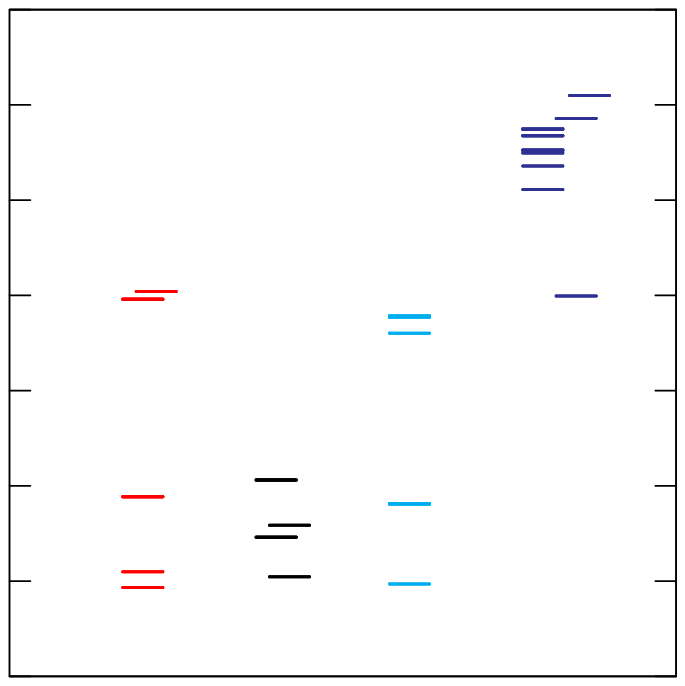}}
   \caption{Sparticle spectrum for mSUGRA parameter set with
   a sneutrino LSP:\;\;$\lamp_{331}(M_{GUT})=0.122$, $\tan
  \beta=10$, $\mzero=100$ GeV, $\mhalf=250$ GeV, $A_0=-100$ GeV,
  $\sgnmu=+1$ and the weak-scale quark mixing is solely in the up sector.
  Numerically the spectrum is given in Table~\ref{spectra}.
  We have omitted $\tilde s_L,\tilde c_L$ which are almost 
 degenerate with $\tilde d_L,\tilde u_L$, respectively.
 We have combined $\tilde u_R,\tilde c_R,\tilde s_R$ into $\tilde q_R$.
 \label{fig:benchIII}}
\end{figure}

For the next set of input parameters, we pick a point from Fig.~\ref{fig:snu} with an
SPS1a-like spectrum but with a tau sneutrino LSP, in order to obtain
different signatures, \textit{i.e.} here we \textit{must} choose a non-zero
$\BLam$: \;\;$\lamp_{331}(M_{GUT})=0.122$, $\tan \beta=10$,
$\mzero=100$ GeV, $\mhalf=250$ GeV, $A_0=-100$ GeV, $\sgnmu=+1$ and
the weak-scale quark mixing is solely in the up sector. The spectrum
is displayed in Fig.~\ref{fig:benchIII}. Numerically it is given in
Table~\ref{spectra}. The mass ordering of the lightest sparticles is
$m_{\stau_1} > m_{\neutralino_1} > m_{{\tilde \nu}_\tau}$. The spectrum is
light and would lead to copious SUSY production at the LHC.

%%%%%%%%BEN
% I swapped the ordering of these two (above and below) - this agrees with the
% ordering of the benchmarks more.

 \mymed

\begin{figure}[ht!]
\scalebox{.8}{\include{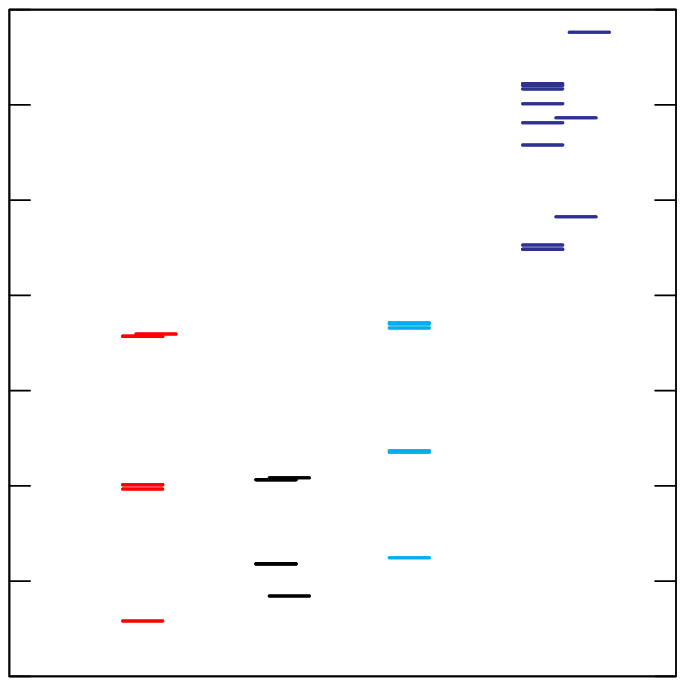}} \caption{Sparticle spectrum for
no-scale mSUGRA parameter set: $\mhalf=600$ GeV, $\tanb=30$, $\sgnmu=+1$,
$\lampp_{212}(M_{GUT})=0.5$.
\label{fig:benchLast}} 
\end{figure}

The final no-scale mSUGRA parameter set consists of a somewhat
heavier spectrum: $M_{1/2}=600$ GeV, $\tan\beta =30$, and $\sgnmu=+1$,
$\lampp_{212}(M_{GUT})=0.5$.
The spectrum is displayed in Fig.~\ref{fig:benchLast}; note the scale
difference to Fig.~\ref{fig:benchI}.
``SUSY'' detection and measurement at this point will be more difficult
than for the no-scale set I, but still easily possible 
\cite{susy-lhc}. We have chosen this point to represent the novel 
mass ordering where the neutralino is the NNNLSP, as discussed in the
previous section: $m_{\neutralino_1} > m_{{\tilde e}_R,{\tilde\mu}_R} >
m_{\stau_R}$.

\subsection{Sparticle Decays}

As pointed out in Ref.~\cite{Allanach:2003eb}, depending upon the
flavour structure of the $\BLam$ couplings, the non-neutralino LSP may
dominantly undergo 2 or 4-body decays. Since we expect two LSPs per
SUSY event, the 4-body decays will significantly change the number of
particles in the final state and therefore alter the potential
signatures. A benchmark where 2-body LSP decays dominate and another
where 4-body LSP decays dominate then seems expedient. We choose a
stau LSP for the 4-body decay, since the calculations for the partial
widths have been completed \cite{ftnt3} and the sneutrino has no
4-body decays. We choose a point with a tau sneutrino LSP, which
necessarily has 2-body decays because of the dominant $\lamp_{331}$
coupling, which it directly couples to.

\mymed

For small $\BLam$, the LSP may be long-lived, leading to sparticles
with a measurable detached decay vertex in the detector.  If the
$\nPsixU$ couplings are even smaller, such decays are longer lived
than $\sim 10^{-6}$ seconds, the staus will decay outside of the
detector and appear as heavily ionising tracks.  This possibility has
been examined recently in the literature~\cite{heavyIonising}, albeit
in a different context. In an intermediate $\nPsixU$ coupling regime,
significant numbers of the staus will decay in the detector providing
heavily ionising tracks with detached vertices. This possibility is
new and interesting and so we shall include a benchmark that covers
it.

\mymed

We note in passing that the case of a long-lived LSP in principle also
include timescales which are relevant for cosmology, \textit{i.e.}
$\tau(\textrm{LSP}) > \mathcal{O}(1\mathrm{sec})$ and the resulting
bounds must be taken into account \cite{Ellis:1990nb}. As we saw, in
the case of R-parity violation, in principle any supersymmetric
particle can be the LSP. We found in our restricted mSUGRA scenarios
either a neutralino, a sneutrino or a stau LSP. As has been recently
investigated in some detail the cosmological bounds on a long-lived
charged particle for example from nucleosynthesis are very different
for charged \cite{Dimopoulos:1988ue,Pospelov:2006sc,Kohri:2006cn,
Kaplinghat:2006qr} and neutral particles \cite{Ellis:1990nb,
Ellis:1984er,Dimopoulos:1988ue,Holtmann:1998gd,Kawasaki:2004qu,
Kanzaki:2006hm,Kanzaki:2006hm,Steffen:2006hw}. For example the
long-lived charged particles can form bound states with the free
nuclei thus affecting their nuclear reactions, which in turn directly
affects the abundance of the light elements \cite{Kohri:2006cn}. A
more detailed analysis is unfortunately beyond the scope of this
paper.

\mymed

$\lampp_{ijk}$ couplings necessarily lead to 4-body stau decays.
These involve many jets in the final state as well as taus from the
$\stau$ decays. Thus the fourth benchmark shall have a non-zero
$\lampp_{ijk}$ at $M_{GUT}$.  More precisely, the four benchmarks
chosen are
\begin{itemize}
\item {\bf BC1}: no-scale mSUGRA with $\lam_{121}(M_{GUT})=0.032$,
  $\tan \beta=13$, $M_{1/2}=400$ GeV, $\sgnmu=+1$.
 The spectrum is shown in Fig.~\protect\ref{fig:benchI}.

\item {\bf BC2}: no-scale mSUGRA with $\lamp_{311}(M_{GUT})=3.5 \times
  10^{-7}$, $\tan \beta=13$, $M_{1/2}=400$ GeV, $\sgnmu=+1$. The
  spectrum is shown in Fig.~\protect\ref{fig:benchI}. The LSP is
  long-lived.

\item {\bf BC3}: mSUGRA with $\lamp_{331}(M_{GUT})=0.122$, $\tan
  \beta=10$, $\mzero=100$ GeV, $\mhalf=250$ GeV, $A_0=-100$ GeV,
  $\sgnmu=+1$ and weak-scale quark mixing solely in the up sector.
  The spectrum is shown in Fig.~\ref{fig:benchIII} and given in
  Table~\ref{spectra}.

\item {\bf BC4}: no-scale mSUGRA with $\lampp_{212}(M_{GUT})=0.5$,
  $\tan \beta=30$, $M_{1/2}=600$ GeV, $\sgnmu=+1$. The spectrum is
%%%% BEN - changed text here to reflect fact that Markus is going to update
%  the figure to the actual spectrum
   shown in Fig.~\protect\ref{fig:benchLast}.
%, except
%  for the $\ssdown_R$ which is 200 GeV lighter, and $\sscharm_R$,
%  $\ssstrange_R$, which are approximately 300 GeV lighter than in the
%  figure.

\end{itemize}

\mymed

In order to calculate the decay rates of sparticles, we pipe the
output of the {\tt SOFTSUSY} spectrum calculation through {\tt
  ISAWIG1.200}.  This is linked to {\tt ISAJET7.64}~\cite{isajet} in
order to calculate the 2-body partial widths of the sparticles and
Higgs'. The output from this procedure is fed into a specially
modified version of the \texttt{HERWIG} program~\cite{herwig} that is
able to calculate partial widths for the 2- and 4-body stau decays
\cite{ftnt4}. We shall now display the decay branching ratios for
sparticles and Higgs'. We omit $h^0$ decays since these follow those
of the Standard Model limit very closely. 

\mymed

%---------------------------------------------------------------%
%--------BC1----------------------------------------------------%
%---------------------------------------------------------------%
\begin{table*}[ht!]
  \centering
\begin{tabular}{cc}
  \begin{tabular}{|l|ll|ll|}
    \hline
    &channel &BR&channel&BR \\
    \hline
$\sstau_1$ & ${\mu^+\bnue e^-\tau^-}$
    &\textbf{0.322}  
&$e^+\bnumu e^-\tau^-$ &\textbf{0.321} \\ 
&$\mu^-\nue e^+ \tau^-$ &\textbf{0.179} 
&$e^-\numu e^+\tau^-$ &\textbf{0.178} \\ \hline
$\neut_1$ &$\sstau_1^+ \tau^-$ &0.498 
&$\sstau_1^- \tau^+$ &0.498 \\
\hline
$\sse_R$ & $e^-\numu$ &\textbf{0.500} 
&$\mu^-\nue$ &\textbf{0.500} \\ \hline
$\ssmu_R$ &$\sstau^+ \mu^- \tau^-$ & 0.512 & $\sstau^- \mu^- \tau^+$ &0.487\\
    \hline  
%%%
$\sstau_2$ &$\neut_1 \tau^-$ &0.630
&$Z^0\sstau_1^-$ &0.176 \\
&$\higgs \sstau_1^-$ &0.194 &&\\ \hline
$\ssnue(\ssnumu)$ &$\neut_1 \nue(\numu)$ &0.924
&$\mu^+(e^+) e^-$ &\textbf{0.075} \\ \hline
$\ssnutau$ &$\neut_1\nutau$ &0.672
&$W^+\sstau_1$ &0.328 \\ \hline
$\sse^-_L(\ssmu^-_L)$ &$\neut_1 e^-(\mu^-)$ &0.919
&$e^-\bnumu(\bnue)$ &\textbf{0.081}\\ \hline
$\ssdown_L(\ssstrange_L)$&$\charge_1^- u(c)$ &0.616 
&$\neut_2 d(s)$ &0.313 \\
&$\charge_2^- u(c)$ &0.038 
&$\neut_1 d(s)$ &0.018 \\
&$\neut_4 d(s)$ &0.014 &&\\ \hline
%&&&$\neut_3 d(s)$ &0.001 \\ 
$\ssup_L(\sscharm_L)$&$\charge_1^+ d(s)$ &0.646 
&$\neut_2 u(c)$ &0.318 \\
&$\charge_2^+ d(s)$ &0.015 
&$\neut_4 u(c)$ &0.011 \\
&$\neut_1 u(c)$ &0.010 &&\\ \hline
%&&&$\neut_3 u(c)$ &0.001 \\ 
%$\ssup_L(\sscharm_L)$ &873.1&$7.80\times10^{-26}$ &$\charge_1^+ d(s)$ &0.641\\
%&&&$\neut_2 u(c)$ &0.315 \\
%&&&$\charge_2^+ d(s)$ &0.019 \\
%&&&$\neut_1 u(c)$ &0.014 \\
%&&&$\neut_4 u(c)$ &0.010 \\
%&&&$\neut_3 u(c)$ &0.001 \\ \hline
$\ssdown_R(\ssstrange_R)$ &$\neut_1 d(s)$ &0.994 
%&$\neut_4 d(s)$ &0.003 
&&\\
%&$\neut_2 d(s)$ &0.003 &&\\
%&&&$\neut_3 d(s)$ &0.001 \\ 
\hline
$\ssup_R(\sscharm_R)$ &$\neut_1 u(c)$ &0.994 
%&$\neut_4 u(c)$ &0.003 
&&\\
%&$\neut_2 u(c)$ &0.003 &&\\
%&&&$\neut_3 u(c)$ &0.001 \\ 
\hline
%$\ssup_R(\sscharm_R)$ &820.9&$3.86\times10^{-25}$ &$\neut_1 u(c)$ &0.993 \\
%&&&$\neut_4 u(c)$ &0.003 \\
%&&&$\neut_2 u(c)$ &0.003 \\
%&&&$\neut_3 u(c)$ &0.001 \\ \hline
 $\ssbottom_1$ & $\charge^-_1 t$&0.360 
 &$\charge^-_2 t$ &0.252 \\
 &$\neut_2 b$ &0.220 
 &$W^- \sstop_1$ &0.120 \\
 &$\neut_1 b$ &0.024 
 &$\neut_3 b$ &0.012 \\
 &$\neut_4 b$ &0.011 &&\\ \hline
$\ssbottom_2$ &$\charge^-_2 t$ &0.408 
&$W^- \sstop_1$ &0.152 \\
&$\charge^-_1 t$ &0.100 
&$\neut_1 b$ &0.127 \\
&$\neut_4 b$ &0.086 
&$\neut_3 b$ &0.067 \\
&$\neut_2 b$ &0.060 &&\\ \hline
%  \end{tabular} 
%\end{table}
%\begin{table}
%  \begin{tabular}{|l|l|l|l|l|}
%    \hline
%     &M(GeV) &T(sec) &channels &BR \\
%    \hline
$\sstop_1$ &$\charge^+_1 b$ &0.440 
&$\neut_1 t$ &0.237 \\
&$\charge^+_2 b$ &0.170 
&$\neut_2 t$ &0.154 \\ \hline
$\sstop_2$ &$\neut_4 t$ &0.235 
&$\charge^+_1 b$ &0.230 \\
&$\charge^+_2 b$ &0.150 
&$Z^0 \sstop_1$ &0.123 \\
&$\neut_3 t$ &0.096 
&$\neut_2 t$ &0.096 \\
&$\higgs t$ &0.057 
&$\neut_1 t$ &0.023 \\ \hline

$\Higgs$ &$b \bar{b}$ &0.854
&$\tau^-\tau^+$ &0.065 \\
&$t \bar{t}$ &0.046 
&$\sstau_2^- \sstau_1^+$ &0.009 \\
&$\sstau_1^- \sstau_2^+$ &0.009 
&$\neut_1 \neut_2$ &0.007 \\

%&$\neut_1 \neut_1$ &0.003 
%&$s \bar{s}$ &0.003 \\
%&$\sstau_1^- \sstau_1^+$ &0.002 &&\\ 
\hline
$\Azero$ &$b \bar{b}$ &0.826
&$\tau^-\tau^+$ &0.063 \\
&$t \bar{t}$ &0.061 
&$\neut_1 \neut_2$ &0.023 \\
%&$\sstau_2^- \sstau_1^+$ &0.004 \\
%&$\sstau_1^- \sstau_2^+$ &0.004 &&\\
%&$s \bar{s}$ &0.003 \\ 
\hline
%&&&$\sstau_1^- \sstau_1^+$ &0.001 \\ \hline
$H^-$ &$\bar{t} b$ &0.853 
&$\bnutau \tau^-$ &0.080 \\
&$\charge_1^- \neut_1$ &0.036 
&$\sstau_1^- \ssbnutau$ &0.028 \\
%&$\bar{c} s$ &0.003&& \\ 
\hline
 \end{tabular} 
%\caption{Sparticle decays for BC1. }
%$\nPsixU$ decays are shown in bold. All
%  decays are prompt.
%
%\label{tab:benchmark decays1}
%\end{table*}
%\begin{table}[ht!]
%  \centering
&
\hspace{1cm}
  \begin{tabular}{|l|ll|ll|}
    \hline
     &channel &BR &channel &BR \\
    \hline
$\neut_2$ &$\ssnutau \bnutau$ &0.091
&$\ssbnutau \nutau$ &0.091 \\
&$\ssnumu \bnumu$ &0.085 
&$\ssbnumu \numu$ &0.085 \\
&$\ssnue \bnue$ &0.085 
&$\ssbnue \nue$ &0.085 \\
&$\sstau_1^- \tau^+$ &0.091
&$\sstau_1^+ \tau^-$ &0.091 \\
&$\ssmu_L^- \mu^+$ &0.045 
&$\ssmu_L^+ \mu^-$ &0.045 \\
&$\sse_L^- e^+$ &0.045 
&$\sse_L^+ e^-$ &0.045 \\
&$\sstau_2^- \tau^+$ &0.031 
&$\sstau_2^+ \tau^-$ &0.031 \\
&$\neut_1 \higgs$ &0.035 &&\\ 
%&$\neut_1 Z$ &0.004 \\
%&$\sse^-_R e^+$ &0.003 
%&$\sse^+_R e^-$ &0.003 \\
%&$\ssmu^-_R \mu^+$ &0.003 
%&$\ssmu^+_R \mu^-$ &0.003 \\ 
\hline
$\neut_3$ &$\charge_1^- W^+$ &0.289
&$\charge_1^+ W^-$ &0.289 \\
&$\neut_2 Z^0$ &0.241
&$\neut_1 Z^0$ &0.102 \\
&$\neut_1 \higgs$ &0.018
&$\sstau_1^- \tau^+$ &0.010 \\
&$\sstau_1^+ \tau^-$ &0.010 
&$\sstau_2^- \tau^+$ &0.008 \\
&$\sstau_2^+ \tau^-$ &0.009 
&$\neut_2 \higgs$ &0.008 \\
%&$\ssnutau \bnutau$ &0.002 
%&$\ssbnutau \nutau$ &0.002 \\
%&$\ssnumu \bnumu$ &0.002 
%&$\ssbnumu \numu$ &0.002 \\
%&$\ssnue \bnue$ &0.002 
%&$\ssbnue \nue$ &0.002 \\ 
\hline
$\neut_4$ &$\charge_1^- W^+$ &0.265 
&$\charge_1^+ W^-$ &0.265 \\
&$\neut_2 \higgs$ &0.175 
&$\neut_1 \higgs$ &0.071 \\
&$\ssnutau \bnutau$ &0.018 
&$\ssbnutau \nutau$ &0.018 \\
&$\ssnumu \bnumu$ &0.018
&$\ssbnumu \numu$ &0.018 \\
&$\ssnue \bnue$ &0.018 
&$\ssbnue \nue$ &0.018 \\
&$\sstau_2^- \tau^+$ &0.017
&$\sstau_2^+ \tau^-$ &0.017 \\
&$\neut_1 Z^0$ &0.018 
&$\neut_2 Z^0$ &0.014 \\
&$\ssmu_L^- \mu^+$ &0.008
&$\ssmu_L^+ \mu^-$ &0.008 \\
&$\sse_L^- e^+$ &0.008 
&$\sse_L^+ e^-$ &0.008 \\
&$\sstau_1^\mp \tau^\pm$ &0.005 &&\\
%&&&$\sstau_1^+ \tau^-$ &0.005 \\
%&$\sse^\pm_R e^\mp$ &0.003 \\
%&&&$\sse^+_R e^-$ &0.003 \\
%&$\ssmu^\pm_R \mu^\mp$ &0.003&& \\
%&&&$\ssmu^+_R \mu^-$ &0.003 \\ 
\hline
$\charge_1^-$ &$\ssnutau \tau^-$ &0.202 
&$\ssnumu \mu^-$ &0.186 \\
&$\ssnue e^-$ &0.186 
&$\sstau_1 \bnutau$ &0.167 \\
&$\ssmu_L \bnumu$ &0.081 
&$\sse_L \bnue$ &0.081 \\
&$\sstau_2 \bnutau$ &0.055
&$\neut_1 W^-$ &0.040 \\ \hline
$\charge_2^-$ &$\neut_2 W^-$ &0.283
&$\charge_1^- Z^0$ &0.253 \\
&$\charge_1^- \higgs$ &0.198 
&$\neut_1 W^-$ &0.081 \\
&$\sstau_2 \bnutau$ &0.044 
&$\ssmu_L \bnumu$ &0.037 \\
&$\sse_L \bnue$ &0.037 
&$\ssbnutau \tau^-$ &0.028 \\
&$\ssbnumu \mu^-$ &0.016 
&$\ssbnue e^-$ &0.016 \\
&$\sstau_1 \bnutau$ &0.006 &&\\ \hline
$\glu$ &$\sstop_1 \bar{t}$ &0.095 
&$\bar{\sstop}_1 t$ &0.095 \\
&$\ssbottom_1 \bar{b}$ &0.077 
&$\bar{\ssbottom}_1 b$ &0.077 \\
&$\ssbottom_2 \bar{b}$ &0.052 
&$\bar{\ssbottom}_2 b$ &0.052 \\
&$\tilde{q} \bar{q}$ &0.250 
&$\bar{\tilde{q}} q$ &0.250 \\
    \hline  
\end{tabular} \\
\end{tabular}
  \caption{Sparticle and Higgs decays for BC1:
  $\lam_{121}(M_{GUT})=0.032$, $\tan \beta=13$, $M_{1/2}=400$ GeV, 
$\mzero=\azero=0$,
  $\sgnmu=+1$. $\nPsixU$ decays are shown in bold font.  Only decays
  with branching ratios greater or equal 0.5\% are shown. All decays
  are prompt. The spectrum is displayed in
  Fig.~\protect\ref{fig:benchI}.
\label{tab:BC1decays}}
\end{table*}

%%%%%%%%%%%%%%%%%%%%%%%%%%%%%%%%%%%%%%%%%%%%%%%%%%%%%%%%%%%%%%%%%%%%%%%%%%%%%%%%%%%%%
%%%%%%%%%%%%%%%%%%%%%%%%%%%%%%%%%%%%%%%%%%%%%%%%%%%%%%%%%%%%%%%%%%%%%%%%%%%%%%%%%%%%%
%%%%%%%%%%%%%%%%%%%%%%%%%%%%%%%%%%%%%%%%%%%%%%%%%%%%%%%%%%%%%%%%%%%%%%%%%%%%%%%%%%%%%

The decays for benchmark point BC1 can be seen in
Table~\ref{tab:BC1decays}. Here and in the following tables of
benchmark points, for decays into the light quark flavours the results
are summed over the flavours $q=u,s,d,c$. We see that, as designed,
the $\stau_1$-LSP is completely dominated by 4-body decays.  Each
4-body decay is accompanied by a tau. Thus good tau identification
would help to identify this scenario.  Furthermore, each 4-body decay
includes a final-state neutrino, resulting in missing transverse
momentum, $\slash \!\!\!p_T$. This should however be reduced compared
to a $\Psix$ model, since the stau momentum is diluted in the 4-body
decay. Also, the left-handed selectron and sneutrinos undergo
significant $\nPsixU$ decays. The heavier neutralinos have significant
branching ratios into charginos as well as $Z^0$-bosons/Higgs and
lighter neutralinos. The number of expected taus from a given SUSY
pair production event is often 4 or more. We also see a difference in
the decays of $\selectron_R$ and $\smuon_R$ due to the presence of the
$\lam_{121}$ coupling. Furthermore, we see from the table that other
$\nPsixU$ couplings that are induced in the RGE running are not large
enough to induce branching ratios greater than 0.005.

%----------------------------------------------------------------------%
%------------BC2-------------------------------------------------------%
%----------------------------------------------------------------------%

\begin{table*}[ht!]
  \centering
\begin{tabular}{cc}
  \begin{tabular}{|l|ll|ll|}
    \hline
     &channel &BR &channel &BR \\
    \hline
$\sstau_1$ &$\bar{u} d$ &\textbf{1.000}&& \\ \hline
$\neut_1$ &$\sstau_1^+ \tau^-$ &0.500 
&$\sstau_1^- \tau^+$ &0.500 \\ \hline
$\sse_R(\ssmu_R)$ &$\sstau_1^+ e^-(\mu^-) \tau^-$ &0.512& 
$\sstau_1^- e^- (\mu^-) \tau^+$ & 0.488\\ \hline
$\sstau_2$ &$\neut_1 \tau^-$ &0.629 
&$Z^0\sstau_1^-$ &0.176 \\
&$\higgs \sstau_1^-$ &0.145 &&\\ \hline
$\ssnue(\ssnumu)$ &$\neut_1 \nue(\numu)$
    &1.000 &&\\ \hline
$\ssnutau$ &$\neut_1\nutau$ &0.672 
&$W^+\sstau_1$ &0.328 \\ \hline
$\sse^-_L(\ssmu^-_L)$ &$\neut_1 e^-(\mu^-)$
     &1.000&&\\ \hline
$\ssdown_L(\ssstrange_L)$ &$\charge_1^- u(c)$ &0.616 
&$\neut_2 d(s)$ &0.313 \\
&$\charge_2^- u(c)$ &0.038 
&$\neut_1 d(s)$ &0.018 \\
&$\neut_4 d(s)$ &0.014 &&\\
%&&&$\neut_3 d(s)$ &0.001 \\ 
\hline
$\ssup_L(\sscharm_L)$ &$\charge_1^+ d(s)$ &0.646 
&$\neut_2 u(c)$ &0.318 \\
&$\charge_2^+ d(s)$ &0.015 
&$\neut_4 u(c)$ &0.011 \\
&$\neut_1 u(c)$ &0.010 &&\\
%&&&$\neut_3 u(c)$ &0.001 \\ 
\hline
$\tilde q_R$ &$\neut_1 q$ &0.994 
&$\neut_4 q$ &0.003 \\
&$\neut_2 q$ &0.002 &&\\
%&&&$\neut_3 d(s)$ &0.001 \\ 
\hline
%$\ssup_R(\sscharm_R)$ &$\neut_1 u(c)$ &0.994 
%&$\neut_4 u(c)$ &0.003 \\
%&$\neut_2 u(c)$ &0.002 &&\\
%&&&$\neut_3 u(c)$ &0.001 \\
% \hline
$\ssbottom_1$ &$\charge^-_1 t$ &0.355 
&$\charge^-_2 t$ &0.252 \\
&$\neut_2 b$ &0.220 
&$W^- \sstop_1$ &0.120 \\
&$\neut_1 b$ &0.024 
&$\neut_3 b$ &0.012 \\
&$\neut_4 b$ &0.011 &&\\ \hline
$\ssbottom_2$ &$\charge^-_2 t$ &0.408 
&$W^- \sstop_1$ &0.152 \\
&$\charge^-_1 t$ &0.100 
&$\neut_1 b$ &0.127 \\
&$\neut_4 b$ &0.086 
&$\neut_3 b$ &0.067 \\
&$\neut_2 b$ &0.060 &&\\ \hline
$\sstop_1$ &$\charge^+_1 b$ &0.440
&$\neut_1 t$ &0.237 \\
&$\charge^+_2 b$ &0.169
&$\neut_2 t$ &0.154 \\ \hline
$\sstop_2$ &$\neut_4 t$ &0.235 
&$\charge^+_1 b$ &0.220 \\
&$\charge^+_2 b$ &0.150 
&$Z^0 \sstop_1$ &0.123 \\
&$\neut_3 t$ &0.096 
&$\neut_2 t$ &0.096 \\
&$\higgs t$ &0.056 
&$\neut_1 t$ &0.023 \\ \hline
$\neut_2$ &$\ssnutau \bnutau$ &0.091 
&$\ssbnutau \nutau$ &0.091 \\
&$\ssnumu \bnumu$ &0.085 
&$\ssbnumu \numu$ &0.085 \\
&$\ssnue \bnue$ &0.085 
&$\ssbnue \nue$ &0.085 \\
&$\sstau_1^- \tau^+$ &0.092 
&$\sstau_1^+ \tau^-$ &0.092 \\
&$\ssmu_L^- \mu^+$ &0.044 
&$\ssmu_L^+ \mu^-$ &0.044 \\
&$\sse_L^- e^+$ &0.044 
&$\sse_L^+ e^-$ &0.044 \\
&$\sstau_2^- \tau^+$ &0.031 
&$\sstau_2^+ \tau^-$ &0.031 \\
&$\neut_1 \higgs$ &0.035 
&& \\  \hline
  \end{tabular} &
\hspace{1cm}
  \begin{tabular}{|l|ll|ll|} 
    \hline
     &channel &BR &channel &BR \\
    \hline
$\neut_3$ &$\charge_1^- W^+$ &0.289 
&$\charge_1^+ W^-$ &0.289 \\
&$\neut_2 Z^0$ &0.241 
&$\neut_1 Z^0$ &0.102 \\
&$\neut_1 \higgs$ &0.018 
&$\sstau_1^- \tau^+$ &0.010 \\
&$\sstau_1^+ \tau^-$ &0.010 
&$\sstau_2^- \tau^+$ &0.008 \\
&$\sstau_2^+ \tau^-$ &0.008 
&$\neut_2 \higgs$ &0.009 \\
\hline
$\neut_4$ &$\charge_1^- W^+$ &0.265
&$\charge_1^+ W^-$ &0.265 \\
&$\neut_2 \higgs$ &0.175
&$\neut_1 \higgs$ &0.072 \\
&$\ssnutau \bnutau$ &0.018 
&$\ssbnutau \nutau$ &0.018 \\
&$\ssnumu \bnumu$ &0.018 
&$\ssbnumu \numu$ &0.018 \\
&$\ssnue \bnue$ &0.018 
&$\ssbnue \nue$ &0.018 \\
&$\sstau_2^- \tau^+$ &0.017 
&$\sstau_2^+ \tau^-$ &0.017 \\
&$\neut_1 Z^0$ &0.018 
&$\neut_2 Z^0$ &0.014 \\
&$\ssmu_L^- \mu^+$ &0.008 
&$\ssmu_L^+ \mu^-$ &0.008 \\
&$\sse_L^- e^+$ &0.008 
&$\sse_L^+ e^-$ &0.008 \\
&$\sstau_1^- \tau^+$ &0.005 
&$\sstau_1^+ \tau^-$ &0.005 \\ \hline
$\charge_1^-$ &$\ssnutau \tau^-$ &0.203 
&$\ssnumu \mu^-$ &0.186 \\
&$\ssnue e^-$ &0.186 
&$\sstau_1 \bnutau$ &0.168 \\
&$\ssmu_L \bnumu$ &0.081 
&$\sse_L \bnue$ &0.081 \\
&$\sstau_2 \bnutau$ &0.060 
&$\neut_1 W^-$ &0.040 \\ \hline
$\charge_2^-$ &$\neut_2 W^-$ &0.283 
&$\charge_1^- Z^0$ &0.253 \\
&$\charge_1^- \higgs$ &0.198 
&$\neut_1 W^-$ &0.082 \\
&$\sstau_2 \bnutau$ &0.044 
&$\ssmu_L \bnumu$ &0.037 \\
&$\sse_L \bnue$ &0.037 
&$\ssbnutau \tau^-$ &0.028 \\
&$\ssbnumu \mu^-$ &0.016 
&$\ssbnue e^-$ &0.016 \\
&$\sstau_1 \bnutau$ &0.006 &&\\ \hline
$\glu$ &$\sstop_1 \bar{t}$ &0.095 
&$\bar{\sstop}_1 t$ &0.095 \\
&$\ssbottom_1 \bar{b}$ &0.077 
&$\bar{\ssbottom}_1 b$ &0.077 \\
&$\ssbottom_2 \bar{b}$ &0.052 
&$\bar{\ssbottom}_2 b$ &0.052 \\
&$\tilde{q} \bar{q}$ &0.250 
&$\bar{\tilde{q}} q$ &0.250 \\
\hline
$\Higgs$ &$b \bar{b}$ &0.854 
&$\tau^-\tau^+$ &0.065 \\
&$t \bar{t}$ &0.046 
&$\neut_1 \neut_2$ &0.007 \\
&$\sstau_2^- \sstau_1^+$ &0.009 
&$\sstau_1^- \sstau_2^+$ &0.009 \\ \hline
$\Azero$ &$b \bar{b}$ &0.826 
&$\tau^-\tau^+$ &0.063 \\
&$t \bar{t}$ &0.061 
&$\neut_1 \neut_2$ &0.023 \\ \hline
$H^-$ &$\bar{t} b$ &0.853 
&$\bnutau \tau^-$ &0.080 \\
&$\charge_1^- \neut_1$ &0.036 
&$\sstau_1^- \ssbnutau$ &0.028 \\ \hline
\end{tabular} \\
\end{tabular}
  \caption{Sparticle and Higgs decays for BC2: $\lamp_{311}(M_{GUT})=3.5
  \times 10^{-7}$, $\tan \beta=13$, $M_{1/2}=400$ GeV, $\mzero=A_0=0$. 
  $\nPsixU$ decays are shown in bold font. Only decays with branching ratios 
  greater than 0.005 are shown. In the tenth row $\tilde q_R=\tilde d_R,
\tilde s_R,\tilde u_R,\tilde c_R$. And the $q$ in the decays is correspondingly
$q=d,s,u,c$.  All decays are prompt except for $\stau_1$, which has a
lifetime of 1.1$\times 10^{-12}$ sec. The spectrum is shown in
Fig.~\protect\ref{fig:benchI}.
\label{tab:BC2decays}}
\end{table*}

%%%%%%%%%%%%%%%%%%%%%%%%%%%%%%%%%%%%%%%%%%%%%%%%%%%%%%%%%%%%%%%%%%%%%%%%%%%%%%%

\mymed

The decays for benchmark point BC2 can be seen in
Table~\ref{tab:BC2decays}. We see that, as designed, the $\stau_1$-LSP
is completely dominated by 2-body decays into non-$b$ jets. Unlike BC1,
the LSP decay comes without a neutrino, meaning that SUSY events do
not necessarily possess the classic missing transverse momentum
$\slash \!\!\!p_T$ signature. A comparison between
Tables~\ref{tab:BC2decays}, and \ref{tab:BC1decays} shows that many of
the decays are identical as a result of the spectrum being
approximately identical. The main differences are in the light
sparticle decays: $\stau_1$, $\selectron_{L,R}$, $\smuon_{L,R}$ and
${\tilde \nu}_{e,\mu}$. The $\lamp_{311}$ coupling is too small to
have a significant affect upon squark decays.  However, in a collider
the main difference will be the existence of detached vertices from
the relatively long-lived $\stau_1$, which has a lifetime of
$\tau_{\tilde\tau_1}=1.1\times 10^{-12}$ secs in its rest-frame. This
corresponds to a rest-frame decay length of $c\tau_{\tilde\tau_1}
\approx 0.3$ mm.  However, when this is boosted into the lab frame
(for example by a factor $\gamma=30$), then the decay-length becomes
about 1 cm. If these $\stau$-LSPs come from other sparticle decays,
once can expect a coincidence of detached vertices with other
particles originating from the interaction point in the detector. Of
course the number of decays will fall approximately exponentially with
radial distance, and so one may obtain decays at radial distances of
$\mu$m-mm, possibly interfering with $b$-tagging. We leave any
in-depth study of such effects to future work using the benchmarks.

\mymed

%---------------------------------------------------------------%
%--------BC3----------------------------------------------------%
%---------------------------------------------------------------%
\begin{table*}[ht!]
 \centering
\begin{tabular}{cc}
 \begin{tabular}{|l|ll|ll|}
   \hline
    &channel &BR &channel &BR \\
   \hline
$\ssnutau$ &$\bar b d$ & \textbf{1.000} && \\ \hline
$\neut_1$ &$\ssbnutau \nutau $ &0.500
&$\ssnutau \bnutau $ &0.500 \\ \hline
$\sstau_1^-$ &  $\nutau b \bar d \tau^-$ & \textbf{0.372}  &  $\bnutau \bar b d \tau^-$ & \textbf{0.372}\\
& $\neut_1 \tau^-$& 0.256 & & \\\hline
$\sstau_2^-$ &$\neut_1 \tau^-$ & 1.000 && \\ \hline
$\ssnue(\ssnumu)$ &$\neut_1 \nue(\numu)$ &0.852 & $\charge_1^+ e^- (\mu^-)$ & 0.107\\
& $\neut_2 \nue (\numu)$ & 0.041 && \\ \hline
$\sse^-_L(\ssmu^-_L)$ &$\neut_1 e^-(\mu^-)$ &0.476&$\charge_1^- \bnue (\bnumu)$&0.331\\
&$\neut_2 e^- (\mu^-)$&0.192&&\\ \hline
$\sse_R^-(\ssmu_R^-)$ &$\neut_1 e^- (\mu^-)$&1.000&& \\[1mm] \hline
$\ssdown_L(\ssstrange_L)$ &$\charge_1^- u(c)$ &0.604
&$\neut_2 d(s)$ &0.307 \\
&$\charge_2^- u(c)$ &0.047
&$\neut_1 d(s)$ &0.024 \\
&$\neut_4 d(s)$ &0.017 && \\\hline
$\ssdown_R$ &$\nu_\tau b$&\textbf{0.494} &$\tau^- t$&\textbf{0.397} \\
&$\neut_1 d$ &0.108
&& \\ \hline
$\ssstrange_R(\tilde u_R,\tilde c_R)\;$ &$\neut_1 s(u,c)$ &0.985
&$\neut_2 s(u,c)$& 0.010\\ \hline
$\tilde u_L$($\tilde c_L$) &$\charge_1^+ d(s)$ &0.650
&$\neut_2 u(c)$& 0.317\\
&$\charge_2^+ d(s)$ &0.015
&$\neut_4 u(c)$& 0.012\\ 
&$\neut_1 u(c)$ &0.006 &&\\ \hline
$\ssbottom_1$ &$\charge^-_1 t$ &0.350
&$\neut_2 b$ &0.275 \\
&$\bnutau d$ & \textbf{0.219}
&$W^- \sstop_1$ & 0.106\\
&$\neut_1 b$ & 0.032
&$\neut_4 b$ & 0.011 \\
&$\neut_3 b$ & 0.007 && \\\hline
$\ssbottom_2$ &$\neut_1 b$ &0.321
&$W^- \sstop_1$ &0.217 \\
&$\bnutau d$ &\textbf{0.125}
&$\neut_4 b$ &0.108 \\
&$\neut_3 b$ &0.093
&$\charge_1^- t$ &0.076 \\
&$\neut_2 b$ &0.060&&\\ \hline
$\sstop_1$ &$\charge^+_1 b$ &0.600
&$\neut_1 t$ &0.157 \\
&$\tau^+ d$ &\textbf{0.124}
&$\neut_2 t$ &0.108 \\
&$\charge_2^+ b$ &0.010
&&\\ \hline
$\sstop_2$ &$\charge^+_2 b$ &0.191
&$\neut_4 t$ &0.185 \\
&$\charge^+_1 b$ &0.176
&$Z^0 \sstop_1$ &0.169 \\
&$\tau^+ d$ &\textbf{0.108}
&$\neut_2 t$ &0.071 \\
&$\neut_3 t$ &0.041
&$\higgs \sstop_1$ &0.033 \\
&$\neut_1 t$ &0.027
&&\\ \hline
$\neut_2$ &$\ssbnutau \nutau$ &0.270
&$\ssnutau \bnutau$ &0.270 \\
&$\sstau_1^+ \tau^-$ &0.219
&$\sstau_1^- \tau^+$&0.219 \\
&$\sstau_2^+ \tau^-$ &0.009
&$\sstau_2^- \tau^+$&0.009 \\ \hline

 \end{tabular}
&
%\label{tab:2benchmark decays1}
%  \caption{BC2, only BR greater than 0.001 channels are shown.}
%
%\end{table}
%\begin{table*}[ht!]
%  \centering
\hspace{1cm}
 \begin{tabular}{|l|ll|ll|}
   \hline
    &channel &BR &channel &BR \\
   \hline
$\neut_3$&$\charge_1^- W^+$ &0.295
&$\charge_1^+ W^-$ &0.295 \\
&$\neut_2 Z^0$ &0.212
&$\neut_1 Z^0$ &0.109 \\
&$\neut_1 \higgs$ &0.020
&$\neut_2 \higgs$ &0.011 \\
&$\sstau_2^- \tau^+$ &0.011
&$\sstau_2^+ \tau^-$ &0.011 \\
&$\ssbnutau \nutau$ &0.005
&$\ssnutau \bnutau$ &0.005 \\ \hline
$\neut_4$ &$\charge_1^- W^+$ &0.252
&$\charge_1^+ W^-$ &0.252 \\
&$\neut_2 \higgs$ &0.127
&$\neut_1 \higgs$ &0.062 \\
&$\ssnutau \bnutau$ &0.037
&$\ssbnutau \nutau$ &0.037 \\
&$\ssnumu \bnumu$ &0.024
&$\ssbnumu \numu$ &0.024 \\
&$\ssnue \bnue$ &0.013
&$\ssbnue \nue$ &0.013 \\
&$\neut_1 Z^0$ &0.020
&$\sstau_2^- \tau^+$ &0.019 \\
&$\sstau_2^+ \tau^-$ &0.019
&$\neut_2 Z^0$ &0.018 \\
&$\ssmu_L^- \mu^+$ &0.009
&$\ssmu_L^+ \mu^-$ &0.009 \\
&$\sse_L^- e^+$ &0.009
&$\sse_L^+ e^-$ &0.009 \\ \hline
$\charge_1^-$ &$\ssnutau \tau^-$ &0.632
&$\sstau_1^- \nutau$ &0.354 \\
&$\sstau_2^- \nutau$ &0.013
&& \\ \hline
$\charge_2^-$ &$\neut_2 W^-$ &0.284
&$\charge_1^- Z^0$ &0.233 \\
&$\charge_1^- \higgs$ &0.161
&$\neut_1 W^-$ &0.062 \\
&$\ssmu_L^- \bnumu$ &0.050  &$\sse_L^- \bnue$ &0.050 \\
&$\sstau_2^- \bnutau$ &0.048
&$\ssbnutau \tau^-$ &0.039 \\
&$\sstau_1^- \bnutau$ &0.035
&$\ssbnumu \mu^-$&0.019 \\
&$\ssbnue e^-$ &0.019
&& \\ \hline
$\glu$ &$\ssbottom_1 \bar b$ &0.114
&$\bar\ssbottom_1 b$ &0.114 \\
&$\sstop_1 \bar{t}$ &0.052
&$\bar{\sstop}_1 t$ &0.052 \\
&$\ssbottom_2 \bar{b}$ &0.049
&$\bar{\ssbottom}_2 b$ &0.049 \\
&$\tilde{q} \bar{q}$ &0.284
&$\bar{\tilde{q}} q$ &0.284 \\ \hline
$\Higgs$ &$b \bar{b}$ &0.787
&$\tau^-\tau^+$ &0.057 \\
&$t \bar{t}$ &0.038
&$\neut_1 \neut_2$ &0.035 \\
&$\charge_1^+ \charge_1^-$ &0.026
&$\sstau_1^- \sstau_1^+$ &0.017 \\
&$\neut_1 \neut_1$ &0.012
&$\neut_2 \neut_2$ &0.010 \\
&$\sstau_2^- \sstau_2^+$ &0.006
&& \\ \hline
$\Azero$ &$b \bar{b}$ &0.661
&$t \bar{t}$ &0.099 \\
&$\neut_1 \neut_2$ &0.067
&$\neut_2 \neut_2$ &0.058 \\
&$\tau^-\tau^+$ &0.048
&$\charge_1^+ \charge_1^-$ &0.027  \\
&$\neut_1 \neut_1$ &0.015
&$\sstau_1^- \sstau_1^+$ &0.010 \\
&$\sstau_1^- \sstau_2^+$ &0.007
&& \\ \hline
$H^-$ &$\bar{t} b$ &0.753
&$\charge_1^- \neut_1$ &0.130 \\
&$\bnutau \tau^-$ &0.073
&$\sstau_1^- \ssbnutau$ &0.021\\
&$\sstau_2^- \ssbnutau$ &0.015
&& \\ \hline
\end{tabular} \\
\end{tabular}
\caption{Sparticle and Higgs decays for BC3:
  $\lamp_{331}(M_{GUT})=0.122$, $\tan \beta=10$, $M_{1/2}=250$ GeV,
  $\mzero=100$ GeV, $A_0=-100$ GeV.  $\nPsixU$ decays are shown in
  bold font. Only decays with branching ratios greater than 0.005 are
  shown. All decays are prompt. 
 \label{tab:BC3decays}}
\end{table*}

\mymed

The decays for benchmark point BC3 are shown in
Table~\ref{tab:BC3decays}. Here the tau sneutrino is the LSP and it
necessarily couples to the dominant $ \nPsixU$ operator.  Therefore it
will always decay via the 2-body mode to the purely hadronic final
state $b \bar d$ ($\bar b d$). One of the daughter jets involves a
bottom quark, providing for the possibility of using $b-$tagging in
order to help identify SUSY events. The neutralino is the NLSP and the
only decay mode open is the 2 body decay to $\tilde\nu_\tau\bar\nu_
\tau$ ($\tilde\nu^*_\tau\nu_\tau$), which results in missing $p_{
\mathrm{T}}$ in the final state. Both $\charge_1^\pm$ and $\neut_2$ have significant
branching ratios to tau leptons/stau sleptons. Thus the cascade
decays of left-handed squarks and indirectly thus also the gluinos
should still lead to significant number of tau's in the final
state. We also see from the table that other $\nPsixU$ couplings that
are induced in the RGE running are not sufficiently large to induce
branching ratios greater than 0.005.

\mymed

%---------------------------------------------------------------%
%--------BC4----------------------------------------------------%
%---------------------------------------------------------------%
\begin{table*}[ht!]
  \centering
\begin{tabular}{cc}
  \begin{tabular}{|l|ll|ll|}
    \hline
     &channel &BR &channel &BR \\
    \hline
$\sstau_1$ &$cds\tau^-$ &\textbf{0.794}
&$\bar{c} \bar{d} \bar{s} \tau^-$ &\textbf{0.206} \\ \hline
$\neut_1$ &$\sstau_1^+ \tau^-$ &0.466 
&$\sstau_1^- \tau^+$ &0.466 \\
&$\ssmu^-_R \mu^+$ &0.017 
&$\ssmu^+_R \mu^-$ &0.017 \\
&$\sse^-_R e^+$ &0.017 
&$\sse^+_R e^-$ &0.017 \\ \hline
$\sstau_2$ &$Z^0\sstau_1^-$ &0.476 
&$\higgs \sstau_1^-$ &0.377 \\
&$\neut_1 \tau^-$ &0.148&& \\ \hline
$\ssnue(\ssnumu)$ &$\neut_1 \nue(\numu)$ &1.000 &&\\ \hline
$\ssnutau$ &$W^+\sstau_1$ &0.885 
&$\neut_1\nutau$ &0.115 \\ \hline
$\sse^-_L(\ssmu^-_L)$ &$\neut_1 e^-(\mu^-)$ &1.000&&\\ \hline
$\sse_R(\ssmu_R)$ &$\sstau_1^+ e^-(\mu^-) \tau^-$ &0.584 
&$\sstau_1^- e^-(\mu^-) \tau^+$ &0.416 \\ \hline
$\ssdown_L(\ssstrange_L)$ &$\charge_1^- u(c)$ &0.629 
&$\neut_2 d(s)$ &0.318 \\
&$\charge_2^- u(c)$ &0.026 
&$\neut_1 d(s)$ &0.016 \\
&$\neut_4 d(s)$ &0.010 && \\ \hline
%&&&$\neut_3 d(s)$ &0.001 \\ \hline
%$\ssup_L(\sscharm_L)$ &$\charge_1^+ d(s)$ &0.642 
%&$\neut_2 u(c)$ &0.317 \\
%&$\charge_2^- d(s)$ &0.017 
%&$\neut_1 u(c)$ &0.012 \\
%&$\neut_4 u(c)$ &0.012 &&\\ \hline
$\ssdown_R(\ssstrange_R)$ &$\bar{c} \bar{s}(\bar{d})$ &\textbf{0.988} 
&$\neut_1 d(s)$ &0.012 \\ \hline
$\ssup_L(\sscharm_L)$ &$\charge_1^+ d(s)$ &0.648 
&$\neut_2 u(c)$ &0.320 \\
&$\neut_1 u(c)$ &0.013 
&$\charge_2^+ d(s)$ &0.011 \\
&$\neut_4 u(c)$ &0.008 && \\ \hline
$\ssup_R$ &$\neut_1 u$ &0.997
&& \\ \hline
$\sscharm_R$ &$\bar{s} \bar{d}$ &\textbf{0.953}
&$\neut_1 c$ &0.047 \\ \hline
%&&&$\neut_2 d(s)$ &0.001 \\ \hline
%$\ssup_R(\sscharm_R)$ &$\neut_1 u(c)$ &0.997 
%&$\neut_4 u(c)$ &0.002 \\
%&&&$\neut_2 u(c)$ &0.001 \\ \hline
$\ssbottom_1$ &$\charge^-_1 t$ &0.367 
&$\neut_2 b$ &0.210 \\
&$\charge^-_2 t$ &0.186 
&$W^- \sstop_1$ &0.077 \\
&$\neut_3 b$ &0.070 
&$\neut_4 b$ &0.049 \\
&$\neut_1 b$ &0.040 && \\ \hline
$\ssbottom_2$ &$\charge^-_2 t$ &0.443 
&$W^- \sstop_1$ &0.133 \\
&$\neut_4 b$ &0.125 
&$\neut_3 b$ &0.109 \\
&$\charge^-_1 t$ &0.103 
&$\neut_2 b$ &0.061 \\
&$\neut_1 b$ &0.026 &&\\ \hline
%  \end{tabular} \label{tab:2benchmark decays}
%  \caption{BC2, only BR greater than 0.001 channels are shown.}
%\end{table}
%\begin{table}[ht!]
%  \centering
%  \begin{tabular}{|l|l|l|l|l|}
%    \hline
%     &M(GeV) &T(sec) &channels &BR \\
%    \hline
$\sstop_1$ &$\charge^+_1 b$ &0.243 
&$\charge^+_2 b$ &0.242 \\
&$\neut_1 t$ &0.210 
&$\neut_3 t$ &0.170 \\
&$\neut_2 t$ &0.100 
&$\neut_4 t$ &0.035 \\ \hline
$\sstop_2$ &$\charge^+_1 b$ &0.239 
&$\neut_4 t$ &0.200 \\
&$\charge^+_2 b$ &0.174 
&$\neut_3 t$ &0.109 \\
&$\neut_2 t$ &0.105 
&$Z^0 \sstop_1$ &0.091 \\
&$\higgs \sstop_1$ &0.061 
&$\neut_1 t$ &0.019 \\ \hline
$\neut_2$ &$\sstau_1^+ \tau^-$ &0.127 
&$\sstau_1^- \tau^+$ &0.127 \\
&$\ssnutau \bnutau$ &0.086 
&$\ssbnutau \nutau$ &0.086 \\
&$\ssnumu \bnumu$ &0.065 
&$\ssbnumu \numu$ &0.065 \\
&$\ssnue \bnue$ &0.064 
&$\ssbnue \nue$ &0.064 \\
&$\ssmu_L^- \mu^+$ &0.054 
&$\ssmu_L^+ \mu^-$ &0.054 \\
&$\sse_L^- e^+$ &0.054 
&$\sse_L^+ e^-$ &0.054 \\
&$\sstau_2^- \tau^+$ &0.041 
&$\sstau_2^+ \tau^-$ &0.041 \\
&$\neut_1 \higgs$ &0.014
&& \\ \hline
%&&&$\neut_1 Z$ &0.001 \\ \hline
  \end{tabular} 
\hspace{1cm}
%\label{tab:2benchmark decays1}
%  \caption{BC2, only BR greater than 0.001 channels are shown.}
%
%\end{table}
%\begin{table*}[ht!]
%  \centering
  \begin{tabular}{|l|ll|ll|}
    \hline
     &channel &BR &channel &BR \\
    \hline
$\neut_3$&$\charge_1^- W^+$ &0.245 
&$\charge_1^+ W^-$ &0.245 \\
&$\neut_2 Z^0$ &0.219 
&$\neut_1 Z^0$ &0.084 \\
&$\sstau_1^- \tau^+$ &0.055 
&$\sstau_1^+ \tau^-$ &0.055 \\
&$\sstau_2^- \tau^+$ &0.033 
&$\sstau_2^+ \tau^-$ &0.033 \\
&$\neut_1 \higgs$ &0.017 
&$\neut_2 \higgs$ &0.008 \\ \hline
$\neut_4$ &$\charge_1^- W^+$ &0.240 
&$\charge_1^+ W^-$ &0.240 \\
&$\neut_2 \higgs$ &0.186 
&$\neut_1 \higgs$ &0.070 \\
&$\sstau_2^- \tau^+$ &0.040 
&$\sstau_2^+ \tau^-$ &0.040 \\
&$\sstau_1^- \tau^+$ &0.037 
&$\sstau_1^+ \tau^-$ &0.037 \\
&$\neut_1 Z^0$ &0.017 
&$\neut_2 Z^0$ &0.011 \\
&$\ssnutau \bnutau$ &0.010 
&$\ssbnutau \nutau$ &0.010 \\
&$\ssnumu \bnumu$ &0.010 
&$\ssbnumu \numu$ &0.010 \\
&$\ssnue \bnue$ &0.010 
&$\ssbnue \nue$ &0.010 \\
&$\ssmu_L^- \mu^+$ &0.005 
&$\ssmu_L^+ \mu^-$ &0.005 \\
&$\sse_L^- e^+$ &0.005 
&$\sse_L^+ e^-$ &0.005 \\
%&&&$\ssmu^-_R \mu^+$ &0.001 \\
%&&&$\ssmu^+_R \mu^-$ &0.001 \\
%&&&$\sse^-_R e^+$ &0.001 \\
%&&&$\sse^+_R e^-$ &0.001 \\ 
    \hline
$\charge_1^-$ &$\sstau_1 \bnutau$ &0.246 
&$\ssnutau \tau^-$ &0.185 \\
&$\ssnumu \mu^-$ &0.136
&$\ssnue e^-$ &0.136 \\
&$\ssmu_L \bnumu$ &0.103 
&$\sse_L \bnue$ &0.103 \\
&$\sstau_2 \bnutau$ &0.076 
&$\neut_1 W^-$ &0.016 \\ \hline
$\charge_2^-$ &$\neut_2 W^-$ &0.247 
&$\charge_1^- Z^0$ &0.232 \\
&$\charge_1^- \higgs$ &0.200 
&$\neut_1 W^-$ &0.085 \\
&$\ssbnutau \tau^-$ &0.070 
&$\sstau_1 \bnutau$ &0.065 \\
&$\sstau_2 \bnutau$ &0.041 
&$\ssmu_L \bnumu$ &0.020 \\
&$\sse_L \bnue$ &0.020 
&$\ssbnumu \mu^-$ &0.009 \\
&$\ssbnue e^-$ &0.009 &&\\ \hline
$\glu$ &$\ssstrange_R \bar{s}$ &0.111
&$\bar{\ssstrange}_R s$ &0.111 \\
&$\ssdown_R \bar{d}$ &0.111
&$\bar{\ssdown}_R d$ &0.111 \\
&$\sscharm_R \bar{c}$ &0.107
&$\bar{\sscharm}_R c$ &0.107 \\
&$\sstop_1 \bar{t}$ &0.049 
&$\bar{\sstop}_1 t$ &0.049 \\
&$\ssbottom_1 \bar{b}$ &0.035 
&$\bar{\ssbottom}_1 b$ &0.035 \\
&$\ssbottom_2 \bar{b}$ &0.025 
&$\bar{\ssbottom}_2 b$ &0.025 \\
&$\tilde{q} \bar{q}$ &0.062 
&$\bar{\tilde{q}} q$ &0.062 \\ \hline
%  \end{tabular} \label{tab:2benchmark decays2}
%  \caption{BC2, only BR greater than 0.001 channels are shown.}%
%
%\end{table}
%
%\begin{table}[ht!]
%  \centering
%  \begin{tabular}{|l|l|l|l|l|}
%    \hline
%     &M(GeV) &T(sec) &channels &BR \\
%    \hline  
%% $\higgs$ &118.6&$1.23\times 10^{-22}$ &$b \bar{b}$ &0.801 \\
%% &&&$\tau^-\tau^+$ &0.048 \\
%% &&&$c \bar{c}$ &0.038 \\
%% &&&$g g$ &0.032 \\
%% &&&$W^- c \bar{s}$ &0.012 \\
%% &&&$W^- u \bar{d}$ &0.012 \\
%% &&&$W^+ \bar{c} s$ &0.012 \\
%% &&&$W^+ \bar{u} d$ &0.012 \\
%% &&&$W^- \tau^+ \nutau$ &0.004 \\
%% &&&$W^- \mu^+ \numu$ &0.004 \\
%% &&&$W^- e^+ \nue$ &0.004 \\
%% &&&$W^+ \tau^- \bnutau$ &0.004 \\
%% &&&$W^+ \mu^- \bnumu$ &0.004 \\
%% &&&$W^+ e^- \bnue$ &0.004 \\
%% &&&$s \bar{s}$ &0.002 \\
%% &&&$\gamma \gamma$ &0.002 \\
%&&&$Z b \bar{b}$ &0.001 \\
%&&&$Z s \bar{s}$ &0.001 \\
%&&&$Z d \bar{d}$ &0.001 \\
%&&&$Z c \bar{c}$ &0.001 \\
%&&&$Z u \bar{u}$ &0.001 \\ 
%\hline
$\Higgs$ &$b \bar{b}$ &0.776 
&$\tau^-\tau^+$ &0.061 \\
&$\sstau_2^- \sstau_1^+$ &0.058 
&$\sstau_1^- \sstau_2^+$ &0.058 \\
&$\sstau_1^- \sstau_1^+$ &0.042 
&& \\ \hline
$\Azero$ &$b \bar{b}$ &0.776 
&$\tau^-\tau^+$ &0.061 \\
&$\sstau_2^- \sstau_1^+$ &0.058 
&$\sstau_1^- \sstau_2^+$ &0.058 \\
&$\sstau_1^- \sstau_1^+$ &0.042 
&& \\ \hline
$H^-$ &$\bar{t} b$ &0.733 
&$\sstau_1^- \ssbnutau$ &0.193 \\
&$\bnutau \tau^-$ &0.071 
&&\\ \hline
%&&&$\ssbnutau \nutau$ &0.001 \\
%&&&$\ssnumu \bnumu$ &0.001 \\
%&&&$\ssbnumu \numu$ &0.001 \\
%&&&$\ssnue \bnue$ &0.001 \\
%&&&$\ssbnue \nue$ &0.001 \\ 
\end{tabular} \\
\end{tabular}
  \caption{Sparticle and Higgs decays for BC4: $\lam''_{212}(M_{GUT})=0.5$,
  $\tan \beta=30$, $M_{1/2}=600$ GeV, $M_0=A_0=0$. $\nPsixU$ decays are shown
  in bold font. 
  Only decays with branching ratios greater than 0.005 are shown. All decays
  are prompt. \label{tab:BC4decays}}
\end{table*}

\mymed

We display the decays for benchmark point BC4 in
Table~\ref{tab:BC4decays}.  Here, the $\stau_1$-LSP decays exclusively
into a $\tau$ and 3 non-$b$ jets, since the decays via a virtual
chargino are strongly suppressed, since 
the tau and the daughter quarks
are dominantly $SU(2)$ singlets.
Thus, this point will provide very
little signal $\slash \!\!\!p_T$. The first two generations of right-handed
squarks undergo dominant $\nPsixU$ decays into two jets, altering many
LHC signatures. Aside from these, the decays are rather similar to
those of BC3, as a comparison with Table~\ref{tab:BC3decays} shows,
although some quantitative differences are present in every channel.
The large number of jets in final states from this model could make
analysis of SUSY events difficult: combinatoric backgrounds are likely
in event reconstruction. Examining specific decay chains involving
leptons can help reduce these~\cite{metal}. The combinatoric
backgrounds problem will become worse for high luminosity running,
where pile-up will increase the number of jets in each event. However,
resonant squark production ought to be possible, providing an
additional empirical handle on the model.

\section{CONCLUSIONS \label{sec:conc}}

We have investigated the spectrum of the $\nPsixU$-MSSM embedded in
supergravity, including indirect constraints on parameter space. We
have found different regions of parameter space with a neutralino, a
stau and also a tau sneutrino LSP. Taking these into account, we have
presented the first set of benchmarks in $\nPsixU$
mSUGRA. All of the points are within current search limits and are
consistent with measurements of precision observables. We have picked
different SUSY breaking scenarios: light and heavy.  The heavier
benchmarks are more difficult to detect at the LHC and so should be
used to see how much is possible to achieve through data analyses in
this difficult scenario. The light benchmarks are designed to enable
high statistics analyses in order to determine what is possible to
divine from LHC data in the optimistic case.

\mymed

We have picked the flavour of the $\nPsixU$ couplings in order to
show-case various features relevant for the experiments. In the first
benchmark, four-body decays of the LSP are expected including
neutrinos, leading to complicated final states and $\slash \!\!\!p_T$.
The second benchmark has vertices detached from the interaction point, where LSPs decay into non-bottom jets. Coincidence with SM particles from higher up the cascades are expected from the interaction region. In the third, 2-body decays of a tau sneutrino LSP into jets (one of them a bottom jet) are expected. The fourth benchmark has LSPs
promptly decaying into 3 jets and a tau. It also has a large
$\nPsixU$ coupling capable of producing significant single-squarks at
hadron colliders such as the LHC.  In order to enable analysis of the
proposed benchmark points, we have provided \texttt{HERWIG} files on
the web at URL address \texttt{http://hepforge.cedar.ac.uk/\~\
  allanach/rpv/}, where they are available for public download.

\section{ACKNOWLEDGEMENTS}

MB and HD thank DAMTP, Cambridge, for warm hospitality offered while
part of this work was performed as well as the SUSY working group at
the Cavendish Laboratory, Cambridge, for discussions. HD also thanks
the Aspen Center for Physics for hospitality while part of this work
was performed and the participants of the LHC workshop for stimulating
discussions on this work. We thank Sebastian Grab for computing some
of the 4-body stau LSP decays. We thank the Theoretical Condensed
Matter Group, Bonn, for readily sharing computation resources, and
Markus Schumacher for useful discussions on Higgs physics. This work
was partially supported by PPARC.
 
\mymed

\clearpage
\appendix

\section{$\mathbf{\nPsixU}$ Contribution to
  $\mathbf{\mathbf{Br}(B_{q_k}\ra e^+_m e^-_l)}$}
\label{app:one}

Here, we generalise the calculation of Ref.~\cite{Jang:1997jy} to
include non-degenerate sparticle masses as well as squark mixing.  The
diagrams of interest consist of two s-channel diagrams mediated by
sneutrinos, and a t-channel diagram mediated by left handed up-type
squarks. Using the convention of \cite{Allanach:2001kg}, we write the
squark mixing as:
\beq
    \tilde{u}_{Lj}=c_j\tilde{u}_{1j}-s_j\tilde{u}_{2j}\,,
\eeq

\noindent where the $c_j \equiv \cos \theta_j^{\tilde u}$, $s_j \equiv \sin
\theta_j^{\tilde u}$, $j=1,2,3$ is a generation index and the states
on the right hand side are mass eigenstates.  For convenience, we
define the following quantities:
\bea 
A^j_{lm}&=&\sum^3_{i=1}\frac{\lam^*_{ilm}\lamp_{ij3}}
{m^2_{{\tilde \nu}_i}}\\
B^j_{lm}&=&\sum^3_{i=1}\frac{\lam^*_{iml}\lamp_{i3j}}
{m^2_{{\tilde \nu}_i}}\\
C^j_{lm}&=&\frac{1}{2}\sum^3_{i=1}\lam^{\prime*}_{lij}\lamp_{mi3}
[\frac{c^2_i}{m^2_{\tilde{u}_{1i}}}+\frac{s^2_i}{m^2_{\tilde{u}_{2i}}}]\\
\kap_k &=& \frac{m_{e_k}}{M_{B_{q_j}}}\,.
\eea

\mymed
 
The final expression for the partial width of the decay 
$B_{q_j}\to e^+_me^-_l$ reads:
  \bea
      \Gamma&=&\frac{f^2_{B_{q_j}}M^3_{B_{q_j}}}{64\pi}\nonumber\\
      && \times\sqrt{1-2(\kap^2_m+\kap^2_l)+(\kap^2_m-\kap^2_l)^2}\nonumber \\
      && \times\big\{\big[(|A^j_{lm}|^2+|B^j_{lm}|^2)(1-\kap^2_l-\kap^2_m)+
\nonumber\\
        &&  4\mathrm{Re}(A^j_{lm}B^{j*}_{lm})\kap_l\kap_m\big]+ \nonumber \\
      && |C^j_{lm}|^2[(\kap^2_l+\kap^2_m)-(\kap^2_l-\kap^2_m)^2]+\nonumber \\
      && 2\mathrm{Re}(A^j_{ml}C^{j*}_{ml})\kap_m(1+\kap^2_l-\kap^2_m)+
\nonumber\\
  	&& 2\mathrm{Re}(B^j_{ml}C^{j*}_{ml})\kap_l(1+\kap^2_m-\kap^2_l)\big\},
  \eea
%% \bea
%%     \Gamma&=&\frac{f^2_{B_{q_j}}M^3_{B_{q_j}}}{64\pi}
%%     \sqrt{1-2(\kap^2_m+\kap^2_l)+(\kap^2_m-\kap^2_l)^2}
%%     \left[
%%       (|A^j_{lm}|^2+|B^j_{lm}|^2)(1-\kap^2_l-\kap^2_m)+
%%         4\mathrm{Re}(A^j_{lm}B^{j*}_{lm})\kap_l\kap_m + \right. \nonumber \\
%%     &&  
%%      |C^j_{lm}|^2[(\kap^2_l+\kap^2_m)-(\kap^2_l-\kap^2_m)^2]+
%%      2\mathrm{Re}(A^j_{ml}C^{j*}_{ml})\kap_m(1+\kap^2_l-\kap^2_m)+ 
%%      \left.
%% 	2\mathrm{Re}(B^j_{ml}C^{j*}_{ml})\kap_l(1+\kap^2_m-\kap^2_l)\right],
%% \eea
 where $f_{B_{q_j}}, M_{B_{q_j}}$ are the decay constant and mass of the
    $B_{q_j}$ meson respectively.

%%%%%%%%%%%%%%%%%%%%%%%%%%%%%%%%%%%%%%%%%%%%%%%%%%%%%%%%%%%%%%%%%%%%%%
%%%%%%%%%%%%%%%%%%%%%%%%%%%%%%%%%%%%%%%%%%%%%%%%%%%%%%%%%%%%%%%%%%%%%%


\begin{thebibliography}{99}
\bibitem{ftnt5} 
{The maximal value $\lam^\prime_{331}= 0.15$ leads to a sneutrino
    mass which has been excluded by LEP.}.
\end{thebibliography}
\end{document}